\documentclass[aps,prb,nobibnotes,reprint,unsortedaddress,floatfix]{revtex4-1}
\usepackage[utf8]{inputenc}
\usepackage[T1]{fontenc}
\usepackage{graphicx}
\usepackage{float}
\usepackage{amsmath, amssymb}
\usepackage{xcolor}
\definecolor{blue}{HTML}{1F77B4}
\definecolor{orange}{HTML}{FF7F0E}
\definecolor{green}{HTML}{2CA02C}
\definecolor{red}{HTML}{D62728}
\definecolor{purple}{HTML}{9467BD}
\definecolor{brown}{HTML}{8C564B}
\definecolor{pink}{HTML}{E377C2}
\definecolor{gray}{HTML}{7F7F7F}
\definecolor{yellow}{HTML}{BCBD22}
\definecolor{cyan}{HTML}{17BECF}
\usepackage{hyperref}
\hypersetup{colorlinks=false}

\begin{document}
\title{Molecular collapse in graphene: sublattice symmetry effect} 

\author{Jing Wang}
\email[]{wangjing@hdu.edu.cn}
\affiliation{School of Electronics and Information, Hangzhou Dianzi University, Hangzhou, Zhejiang Province, China}
\affiliation{Departement Fysica, Universiteit Antwerpen, Groenenborgerlaan 171, B-2020 Antwerpen, Belgium}
\affiliation{NANOlab Center of Excellence, University of Antwerp, Belgium}

\author{Mi{\v s}a An\dj{}elkovi\'c}
\affiliation{Departement Fysica, Universiteit Antwerpen, Groenenborgerlaan 171, B-2020 Antwerpen, Belgium}
\affiliation{NANOlab Center of Excellence, University of Antwerp, Belgium}

\author{Gaofeng Wang}
\email[]{gaofeng@hdu.edu.cn}
\affiliation{School of Electronics and Information, Hangzhou Dianzi University, Hangzhou, Zhejiang Province, China}

\author{Francois M. Peeters}
\email[]{francois.peeters@uantwerpen.be}
\affiliation{Departement Fysica, Universiteit Antwerpen, Groenenborgerlaan 171, B-2020 Antwerpen, Belgium}
\affiliation{NANOlab Center of Excellence, University of Antwerp, Belgium}

\date{\today}

\begin{abstract}

    Atomic collapse can be observed in graphene because of its large “effective” fine structure constant, which enables this phenomenon to occur for an impurity charge as low as $Z_c\sim 1-2$. Here, we investigate the effect of the sublattice symmetry on molecular collapse in two spatially separated charge tunable vacancies, that are located on the same (A-A type) or different (A-B type) sublattices. We find that the broken sublattice symmetry: (1) does not affect the location of the main bonding and anti-bonding molecular collapse peaks, (2) but shifts the position of the satellite peaks, because they are a consequence of the breaking of the local sublattice symmetry, and (3) there are vacancy characteristic collapse peaks that only occur for A-B type vacancies, which can be employed to distinguish them experimentally from the A-A type. As the charge, energy, and separation distance increase, the additional collapse features merge with the main molecular collapse peaks. We show that the spatial distribution around the vacancy site of the collapse states allows us to differentiate the molecular from the frustrated collapse.

\end{abstract}


\maketitle 

\section{Introduction}

    Atomic collapse is an electronic phenomenon from quantum electrodynamics (QED) that was predicted to occur in super heavy atomic nuclei~\cite{ref1} nearly 70 years ago. An electron in a Coulomb potential formed by an atomic nucleus occupies circular orbits around the nucleus. As the atomic number $Z$ increases, relativistic quantum effects become more important. Beyond a critical value of $Z_c\approx 170$, the strong Coulomb potential pulls the electron orbit into the positron continuum, which makes the atom unstable. Initially the electron spirals down into the nucleus, after which it spirals away, emitting a positron in a process referred to as atomic collapse. Since the largest known natural element has $Z$ much smaller than the critical value $Z_c$, it has been impossible to observe this phenomenon in real atoms~\cite{ref2, ref3}. 

    However, graphene with Dirac-like charge carriers provides an excellent opportunity for the realization of atomic collapse in two dimensions~\cite{ref4, ref5, ref6}. The condition for $Z > 1/\alpha$ is easier to fulfill in graphene than in “natural” QED systems due to the favorable scaling of the effective “fine structure constant”. In graphene the Fermi velocity $v_F\approx 10^6$  m/s acts as the velocity of light and the effective fine structure constant becomes $\alpha=e^2/(\hbar v_F \kappa)\approx 2.19/\kappa$, where the effective dielectric constant $\kappa$ is largely determined by the environment, and typically takes values between 3 and 10~\cite{ref7, ref8, ref9}. Hence, a charge impurity with a critical charge as low as $Z_c\sim 1-2$ could induce atomic collapse in graphene. This was initially observed by clustering ionized Ca dimers on top of graphene in order to construct a supercritical “artificial nucleus”~\cite{ref10}. Later, researchers reported this phenomenon in other two dimensional graphene systems: (1) a vacancy charged with a scanning tunneling microscope (STM) tip~\cite{ref11} and (2) a STM tip created potential~\cite{ref12}. Various theoretical works have investigated the physical behavior before and after atomic collapse in graphene with an atomic impurity placed on top of the graphene surface~\cite{ref8, ref9, ref13, ref14, ref15, ref16, ref17, ref18, ref19}.

    Recently, atomic collapse studies were extended to the case where two identical impurity charges are put at a certain distance, which can lead to molecular collapse patterns that depend on the distance between them~\cite{ref20}, while opposite charges resemble a dipole like system~\cite{ref21, ref22, ref23, ref24}. A new physical regime termed “frustrated supercritical collapse” was demonstrated~\cite{ref20, ref25}, in which the individual charges placed above the graphene layer are subcritical, while the global system behaves supercritically. In addition, it was shown theoretically~\cite{ref26}, and confirmed experimentally~\cite{ref11}, that a vacancy in graphene can stably host a positive charge, which can be gradually built up by applying voltage pulses with an STM tip~\cite{ref11}. Thus, a vacancy in graphene can act as an artificial nucleus~\cite{ref27}, which furthermore enhances the effective Coulomb potential as compared to a charge impurity placed on top of graphene~\cite{ref10}, because now the charge center resides in the plane of graphene. 

    With improvements in sample quality vacancies are not ubiquitous type of defects. Still they can be purposely formed by removing an atom by ion irradiation~\cite{ref28}. Such a vacancy leads to a quasilocalized zero energy state around the Dirac point as long as the local sublattice symmetry is broken. The corresponding wavefunction around the defect is located only on the sublattice with the majority of atoms. In contrast, when the local sublattice symmetry is preserved no vacancy peak occurs~\cite{ref29, ref30}. Here, we will consider different arrangements of the two vacancy defects and examine how the global sublattice symmetry affects the molecular collapse resonances. By removing atoms of the same or different sublattices, the global sublattice symmetry is broken or preserved, which is reflected in different collapse resonances. The amount of charge on each vacancy and the distance between them $R$ are tuning parameters that allows us to drive the system from subcritical to supercritical. 

    The paper is structured as follows. In section~\ref{sec:model}, we present the model and the method used to obtain the relevant quantities. The main results are discussed in section~\ref{sec:results}. We conclude the paper in section~\ref{sec:concl}.

\section{Model}\label{sec:model}

    In order to simulate vacancies, a discrete model, such as the tight-binding approach, is more suitable than the continuum model based on the Dirac equation. The tight-binding Hamiltonian of graphene is given by the following expression 
    \begin{equation}
        \hat{H} = \sum_{\langle i,j\rangle}(t_{ij}  \hat{a}_i^\dagger \hat{b}_j + H.c.) + \sum_i V_i  \hat{a}_i^\dagger \hat{a}_i + \sum_i V_i  \hat{b}_i^\dagger \hat{b}_i,
    \end{equation}
    where $\hat{a}_i (\hat{b}_i)$ represents the electron creation operator and $\hat{a}_i^\dagger (\hat{b}_i^\dagger)$ is the annihilation operator of an electron at the sublattice A(B) site $i$. $t_{ij}=-2.8$ eV is the hopping strength between the nearest neighbors. The last two terms denote the electrostatic potential $V_i$ felt by the electron in graphene at the site $i$. The Coulomb potential at position $(x,y)$ away from a charged vacancy centered at $(x_0, y_0)$ is 
    \begin{equation}
        V(x-x_0,y-y_0 )=-\hbar v_F \frac{\beta}{ \sqrt{(x-x_0)^2+(y-y_0)^2}},
    \end{equation} 
    where $\beta=Z\alpha$ represents the effective charge. A cutoff at the distance $r^*$ is introduced as an analogue to the size of the nucleus in QED. The natural minimal cutoff length of graphene is the distance between two carbon atoms $a_{cc} = 0.142$ nm. However, the vacancy charge is distributed over several C-atoms around the vacancy, and it was previously found that $r^* = 0.5$ nm is able to explain the experimental atomic collapse results~\cite{ref11}. 

\begin{figure}[htb!]
        \centering
        \includegraphics[width=0.75\columnwidth]{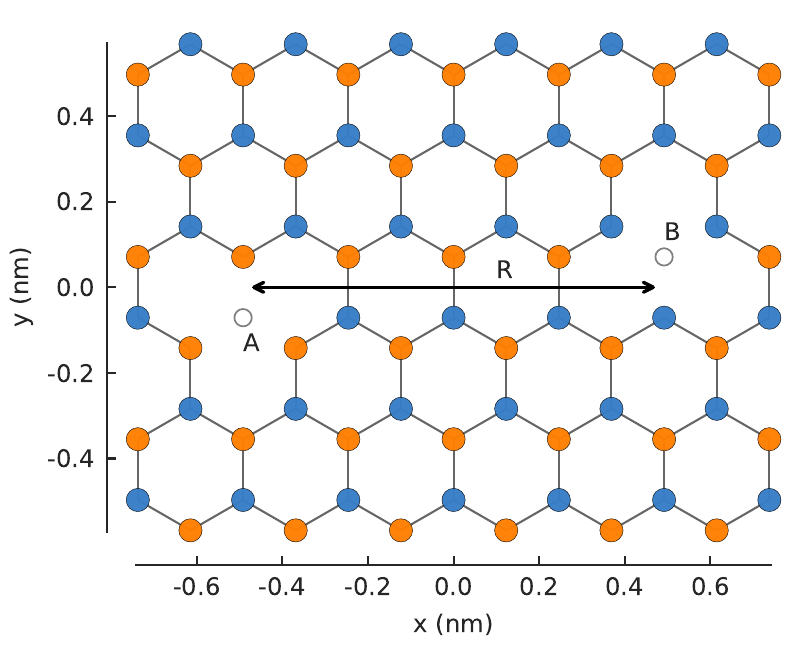}
        \caption{\label{fig:fig1} Schematic representation of two charged vacancies, indicated by gray circles and separated by a distance $R$. The left (right) vacancy is obtained by removing a A (B)-sublattice atom.}
    \end{figure}

    The system we are simulating is illustrated in Fig.~\ref{fig:fig1}, with the two single-atom vacancies placed at a distance $R$. The vacancies can be charged by an STM tip as demonstrated in Ref.~\cite{ref11} and the Coulomb charge centers are situated at the center of each missing atom. Blue and orange circles represent A and B-sublattice atoms, respectively. Global sublattice symmetry is broken (kept) if atoms of the same (different) sublattice are removed. Fig.~\ref{fig:fig1} shows the latter case in an arrangement which we named A-B type vacancy. In contrast A-A type vacancy with two missing A-atoms breaks the global sublattice symmetry. 

    A hexagonal flake with 200 nm sides is used in our numerical calculations, which involves four million carbon atoms. The size of the graphene sheet is large enough such that edge effects are negligible when performing measurements at the center of the system. Furthermore, we modeled a hexagonal flake with armchair edges in order to remove the zero energy edge states. The numerical results were performed using the open source tight-binding package Pybinding~\cite{ref31}, which employs the kernel polynomial expansion for computing the local density of states (LDOS). An energy broadening of  5 meV is used to smooth effects due to the discreteness of the graphene lattice.

\section{Results and Discussion}\label{sec:results}

\begin{figure}[htb!]
        \centering
        \includegraphics[width=0.75\columnwidth]{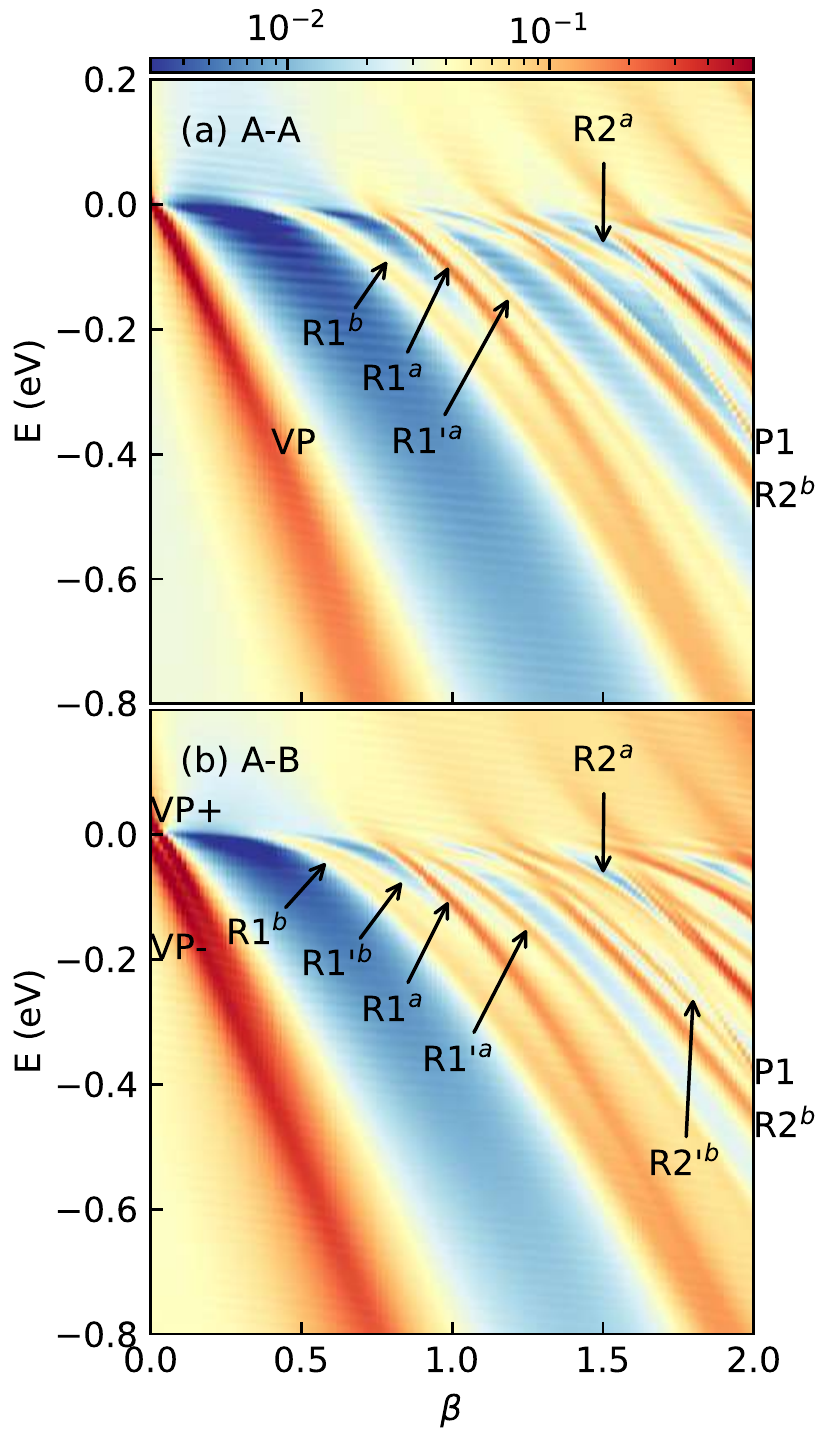}
        \caption{\label{fig:fig2} Contour plot of the LDOS at the edge of one of the vacancies as a function of effective charge and energy. The separation between the two vacancies is $R\approx 5$ nm. (a) A-A type vacancy: the two vacancies are obtained by removing two A-sublattice atoms; (b) A-B type vacancy: an atom from A and from B-sublattice are removed.}
    \end{figure}

    In Fig.~\ref{fig:fig2}, we show a contour plot of the LDOS taken at the edge of one of the vacancies as a function of effective charge and energy for two equally charged vacancies separated by $R\approx 5$ nm of A-A and A-B type. Comparing Fig.~\ref{fig:fig2}(a) with Fig.~\ref{fig:fig2}(b) shows that the global sublattice symmetry breaking has a clear effect on the hole side of the spectrum (negative energy region) where the atomic collapse peaks can be observed, while on the electron side (positive energy region) the LDOS of both systems is the same. 
    
\begin{figure}[htb!]
        \centering
        \includegraphics[width=\columnwidth]{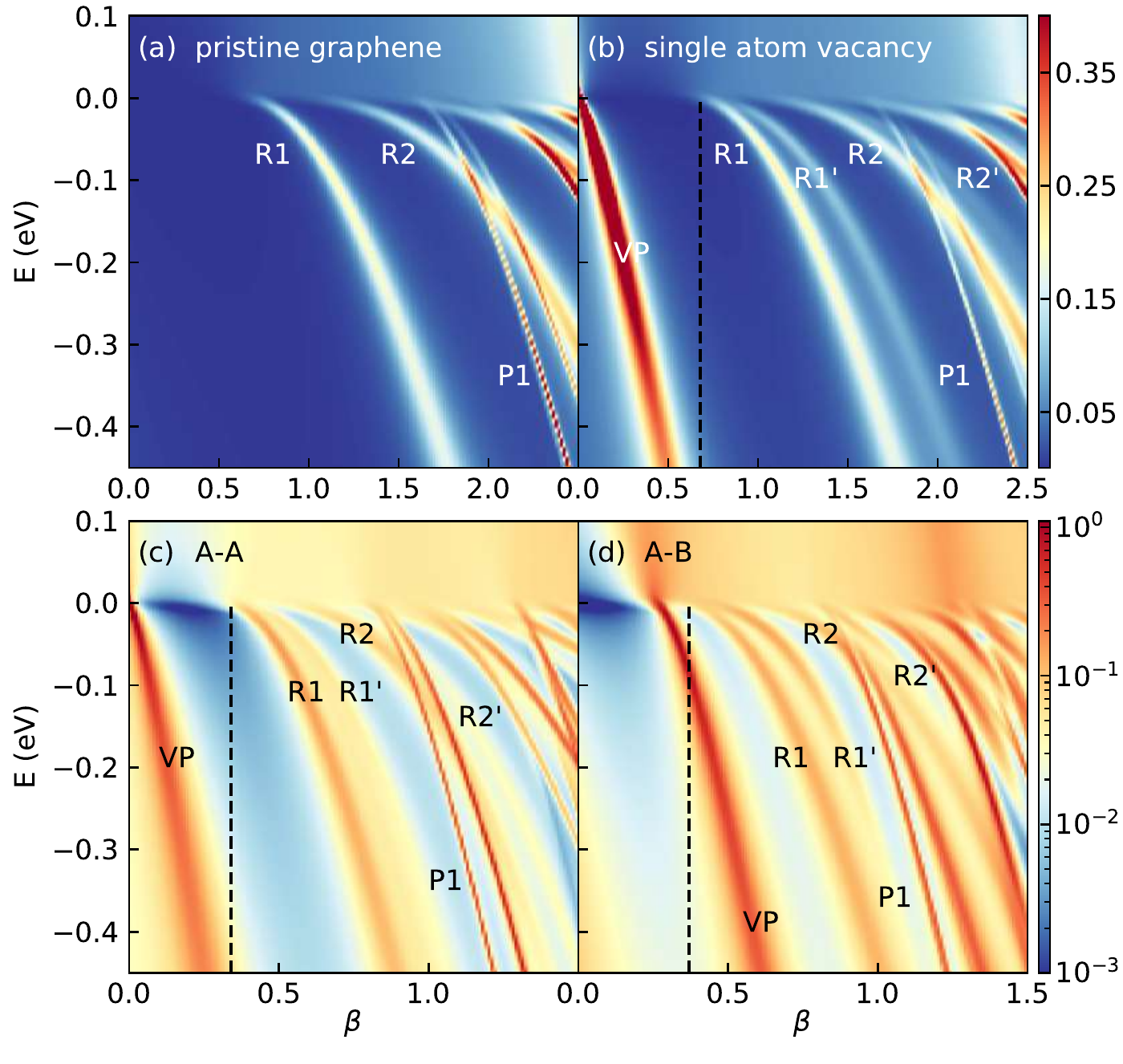}
        \caption{\label{fig:fig3} Contour plot of the LDOS (top: Linear scale; bottom: Log scale) as a function of effective charge and energy, which shows collapse features in (a) pristine graphene with a charge impurity on top of the layer; (b) graphene with a vacancy in which only one atom is removed; (c) A-A type vacancy, $R=\sqrt{3} a_{cc}$, which is the shortest separation between the two sublattices of the same type, and (d) A-B type vacancy, $R=2a_{cc}$.}
    \end{figure}

    In order to understand these results and distinguish the features that are labeled in Fig.~\ref{fig:fig2} we start from a simpler case and show in Fig.~\ref{fig:fig3} the LDOS for (a) pristine graphene with a single charge impurity on top of the layer, (b) single vacancy with one removed atom, the bi-vacancies (c) A-A type with $R= \sqrt{3} a_{cc}$, which is the next-nearest neighbor distance and corresponds with the shortest length between the two sublattices of the same type, and (d) A-B type with $R=2a_{cc}$, or the next-next nearest neighbor sites. By comparing Fig.~\ref{fig:fig3}(a) with Figs.~\ref{fig:fig3}(b-d), we find that (1) a vacancy will introduce a weak satellite R1$^{'}$ (R2$^{'}$) resonance beside the R1 (R2) resonance~\cite{ref11}; (2) The bi-vacancy systems reach supercritical charge for lower values of $\beta$ which is indicated with the vertical dashed lines in Figs.~\ref{fig:fig3}(b, c, and d). R1 and R2 states have the 1s and 2s atomic orbital symmetry~\cite{ref11}, while P1 corresponds to the 2p atomic orbital with angular momentum $m = 1$~\cite{ref1, ref32}. In the case of two sufficiently separated charges, the interaction between the 1s atomic orbitals of the two impurities splits the atomic states into a pair of bonding and anti-bonding molecular orbital states, where the bonding molecular orbital has a lower energy~\cite{ref20}. Taking this into account, we can further distinguish the collapse peaks in Fig.~\ref{fig:fig2}, where the first LDOS resonance peaks R1$^{b(a)}$ show the molecular collapse of the bonding (anti-bonding) molecular orbital. The splitting between these two states vanishes with increase of the effective charge $\beta$ because of a decrease in the overlap between the wavefunctions that become more localized around the separate vacancies. 

    In addition to the molecular collapse peaks, the first strong peak in Figs.~\ref{fig:fig2}(a, b) is the localized vacancy peak (VP) with energy around the Dirac point. According to Lieb’s theorem graphene with equal missing atoms in each sublattice preserves the sublattice symmetry and should not show the zero energy state. Here, in Fig.~\ref{fig:fig2}(b), the same number $N_A$ and $N_B$ of atoms is removed, but nevertheless the LDOS shows two high intensity peaks at $\beta = 0$, which are located on each side of the Dirac point (VP- and VP+). These peaks mimic the bonding and the anti-bonding VP molecular states, as the individual vacancy induced wavefunctions spatially overlap due to the small distance between the defects. We found that these two peaks merge into a zero energy peak if the two vacancies are separated by a large distance and the overlap between the two states is negligible, or if a large broadening parameter is chosen, that would correspond to highly disordered samples. The reason why Lieb’s theorem does not hold lies in the fact that we are dealing with a finite size flake. 
    
\begin{figure}[htb!]
        \centering
        \includegraphics[width=\columnwidth]{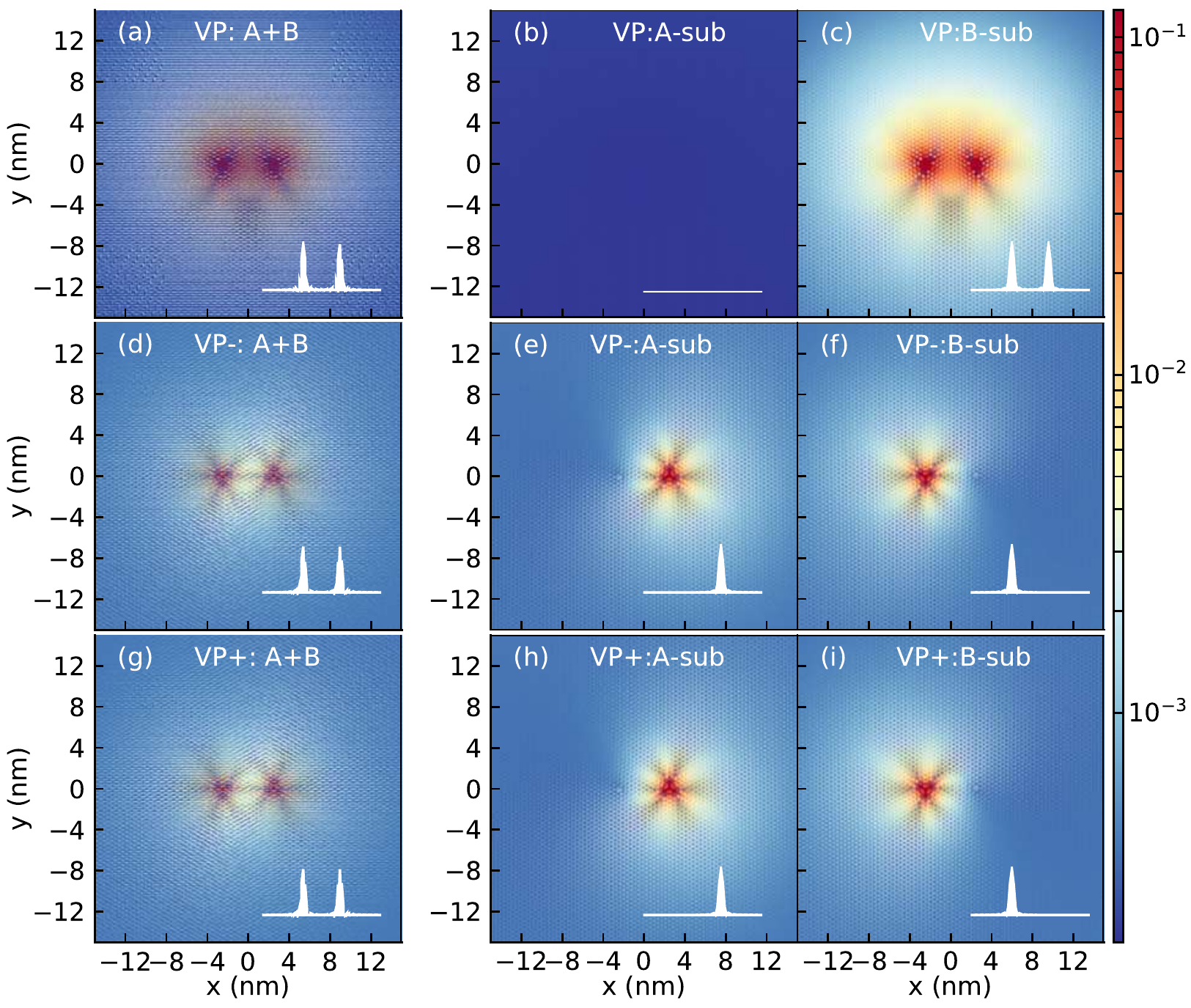}
        \caption{\label{fig:fig4} Spatial LDOS of the vacancy peak at $\beta=0$ for (a-c) A-A type vacancy ($E = 0$ eV) and (d-i) A-B type vacancy, above (VP+, $E = 0.019~\text{eV}$) and (VP-, $E = -0.019~\text{eV}$) below the Dirac point. VP+ and VP- peaks are only mimicing the bonding and the anti-bonding VP molecular states, because their energy splits but the spatial LDOS remains the same. The left column gives the total LDOS while the middle (right) column shows the spatial LDOS on the A(B) sublattice. The insets show cuts of the LDOS along the x-axis for $y=\pm a_{cc}/2$ (left column), $y=-a_{cc}/2$ (middle column) and $y=+a_{cc}/2$ (right column). The separation distance between the vacancies is $R\approx 5$ nm.}
    \end{figure}

    In Fig.~\ref{fig:fig4} we show the LDOS (left figures) corresponding to the vacancy peaks and we show separately the LDOS on the A and B sublattice (right figures). From Fig.~\ref{fig:fig4}(a-c) we notice that the VP of the A-A type defect is fully localized on the B-sublattice, similar to what was previously demonstrated in Refs.~\cite{ref11, ref27} for the single vacancy problem. For both vacancy peaks of the A-B type (Figs.~\ref{fig:fig4}(d-i)) we find that the LDOS of the VP is equally split between the A and B sublattices. The vacancy of A(B)-sublattice has only a nonzero LDOS on the B(A)-sublattice.

\begin{figure}[htb!]
        \centering
        \includegraphics[width=0.75\columnwidth]{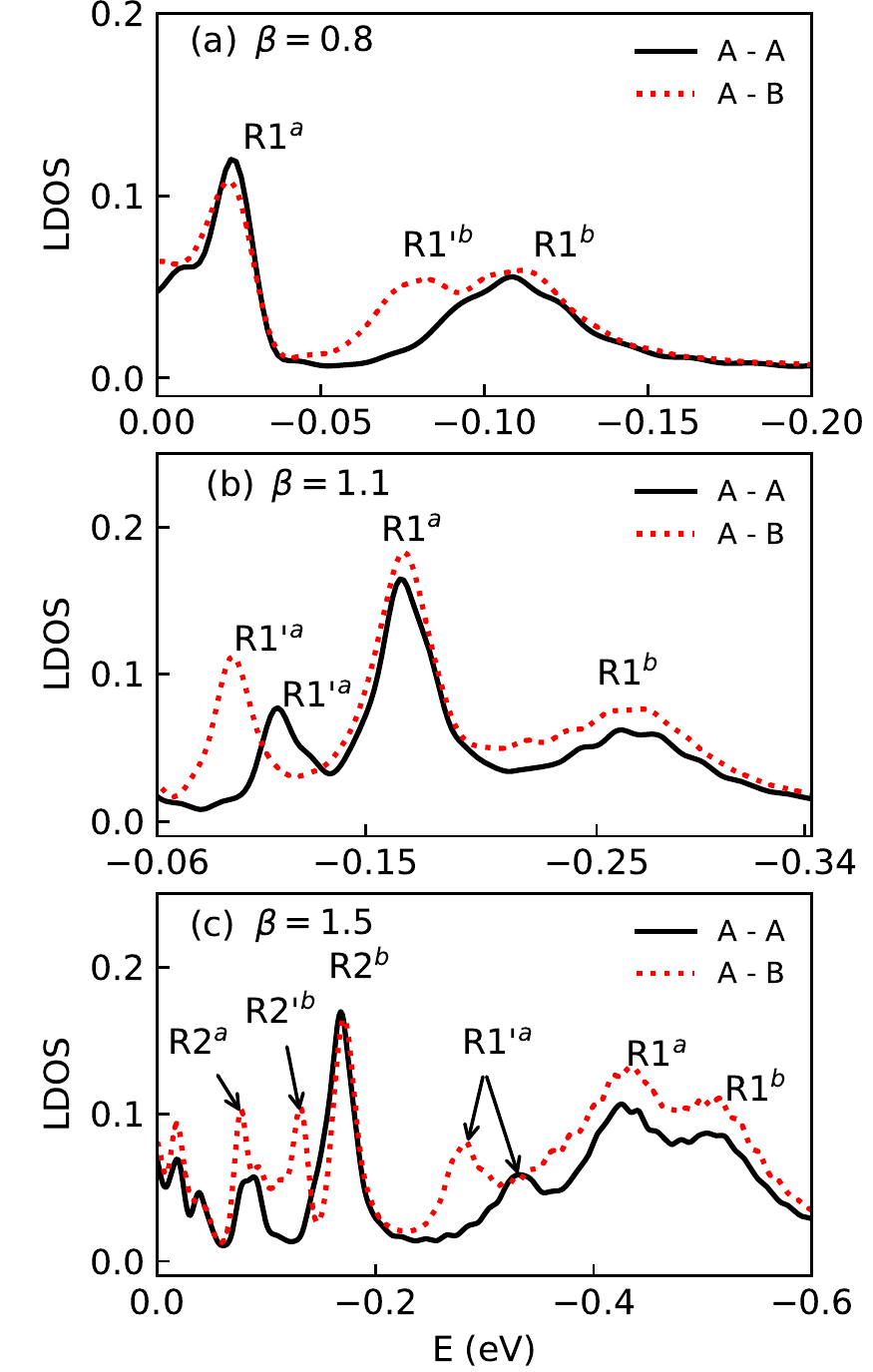}
        \caption{\label{fig:fig5} The LDOS cuts of A-A and A-B type vacancies for (a) $\beta=0.8$, (b) $\beta=1.1$ and $\beta=1.5$, obtained from Fig.~\ref{fig:fig2}.}
    \end{figure}

    The A-B type two vacancy system has additional LDOS peaks that are not present in the A-A type system which can be more clearly distinguished in Fig.~\ref{fig:fig5} where we show the LDOS for three different values of $\beta$. The main bonding and anti-bonding molecular collapse peaks (R1$^b$, R1$^a$, R2$^b$ and R2$^a$) of A-A type and A-B type vacancy have the same energy. The additional satellite collapse peaks (R1'$^{b}$, R1'$^{a}$ and R2'$^{b}$) originate from the presence of the vacancies and are a consequence of the breaking of the local sublattice symmetry as shown in Ref.~\cite{ref11}. They show a much weaker LDOS signature and fade or merge quickly into the main molecular collapse peaks as the charge increases. Energy at which the R1'$^{a}$ peak appears differs in the two cases, while R1'$^{b}$ and R2'$^{b}$ peaks, as noted earlier, are additional peaks that are present only in the case of A-B type vacancy.

\begin{figure}[htb!]
        \centering
        \includegraphics[width=\columnwidth]{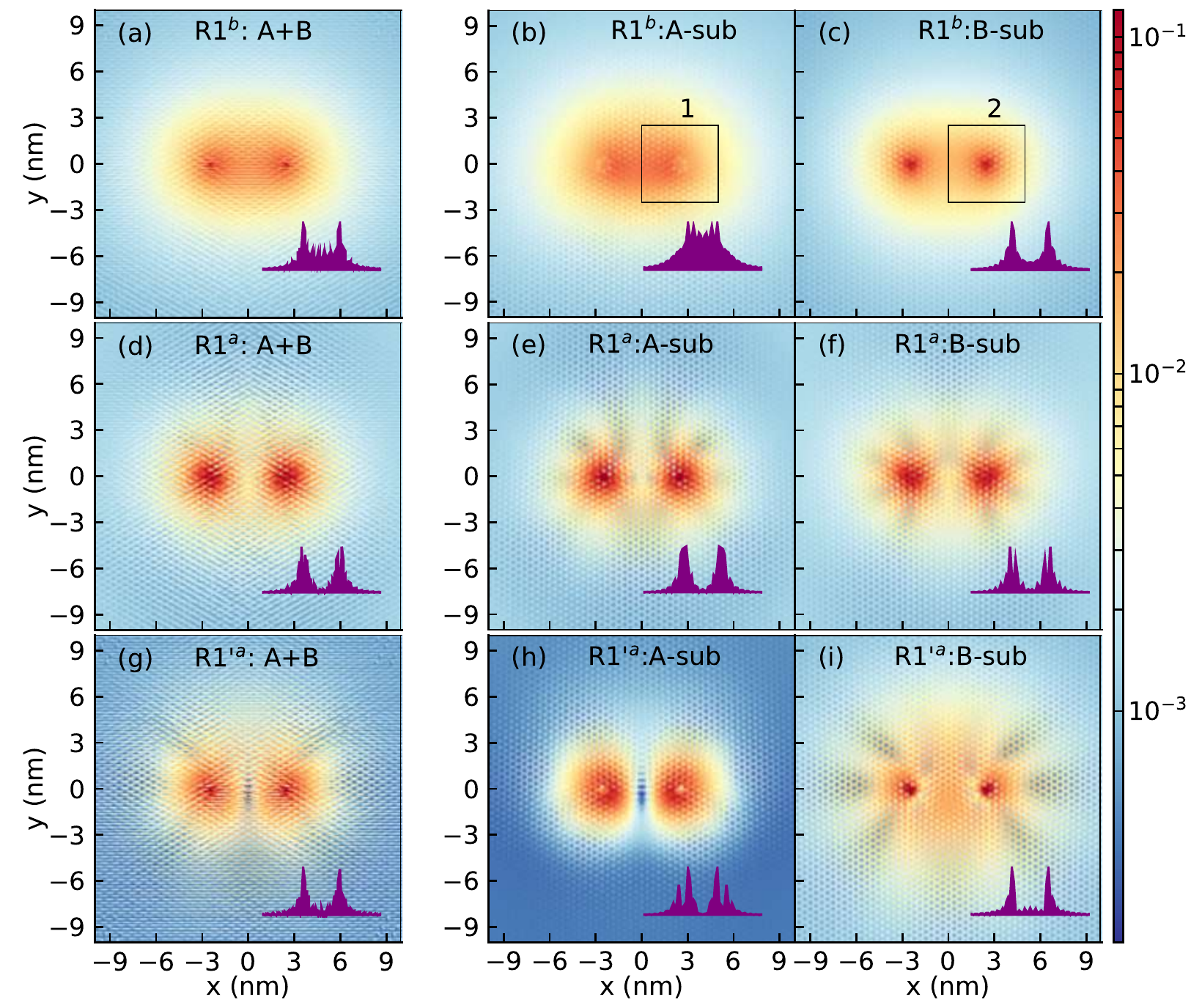}
        \caption{\label{fig:fig6} Spatial LDOS of A-A type vacancies for different resonances labeled in Fig.~\ref{fig:fig2}(a), $(E,\beta) = (-0.1~\text{eV},~0.8)$, $(-0.16~\text{eV},~1.1)$, $(-0.11~\text{eV},~1.1)$ from top to bottom, respectively. Left column shows the total LDOS and the sublattice components are plotted in the middle (A) and the right (B) columns. The insets show cuts of the LDOS along the x-axis for $y=\pm a_{cc}/2$ (left column), $y=-a_{cc}/2$ (middle column) and $y=+a_{cc}/2$ (right column). The separation distance between the two vacancies is $R\approx 5$ nm.}
    \end{figure}

\begin{figure}[htb!]
        \centering
        \includegraphics[width=\columnwidth]{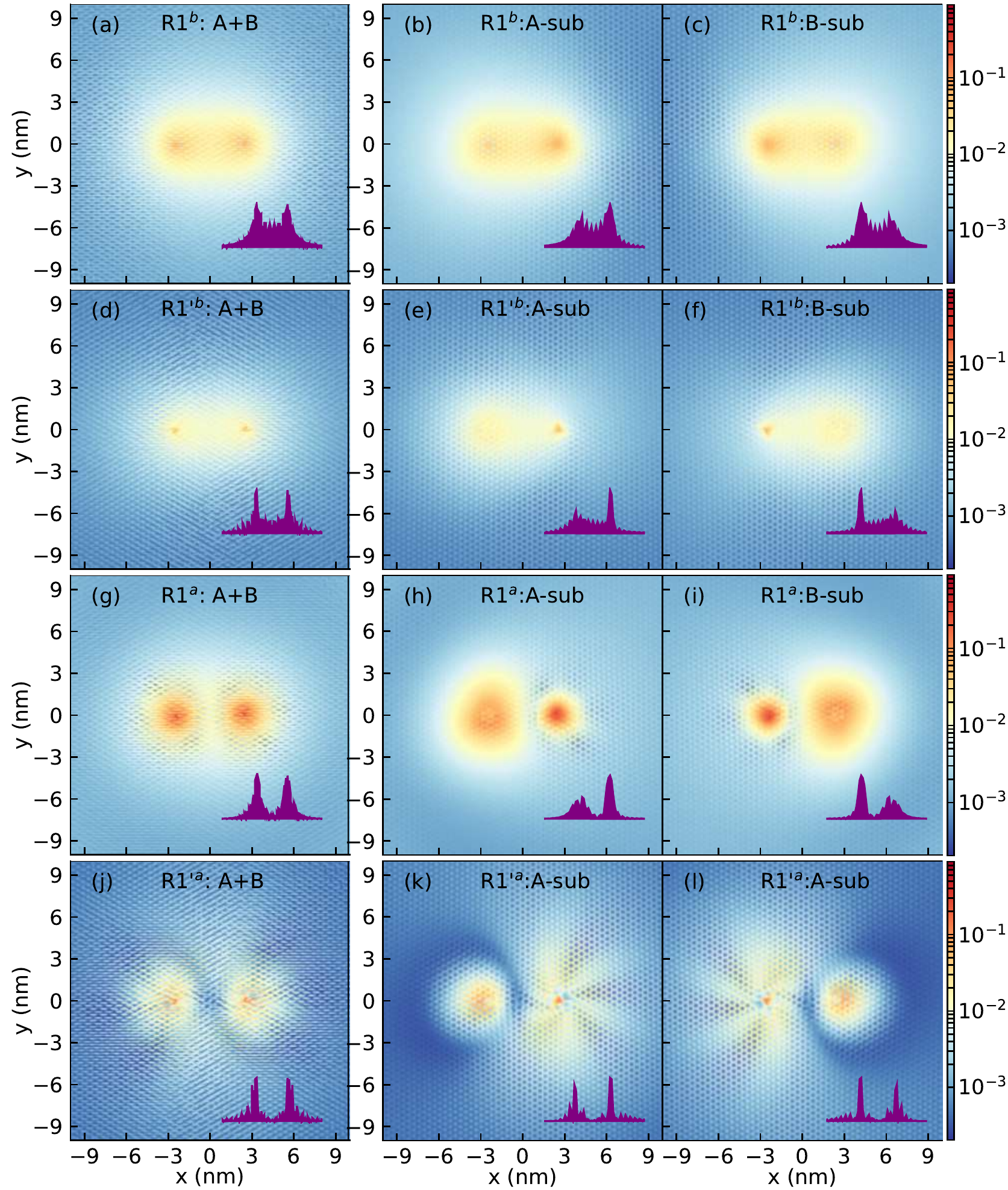}
        \caption{\label{fig:fig7} Spatial LDOS as in Fig.~\ref{fig:fig6}, but now for the A-B type vacancies with $R\approx 5$ nm. Different resonances are labeled in Fig.~\ref{fig:fig2}(b), and we show results for $(E,\beta) = (-0.11~\text{eV},~0.8)$, $(-0.085~\text{eV},~0.8)$, $(-0.16~\text{eV},~1.1)$ and $(-0.089~\text{eV},~1.1)$ from top to bottom, respectively.}
    \end{figure}

\begin{figure}[htb!]
        \centering
        \includegraphics[width=\columnwidth]{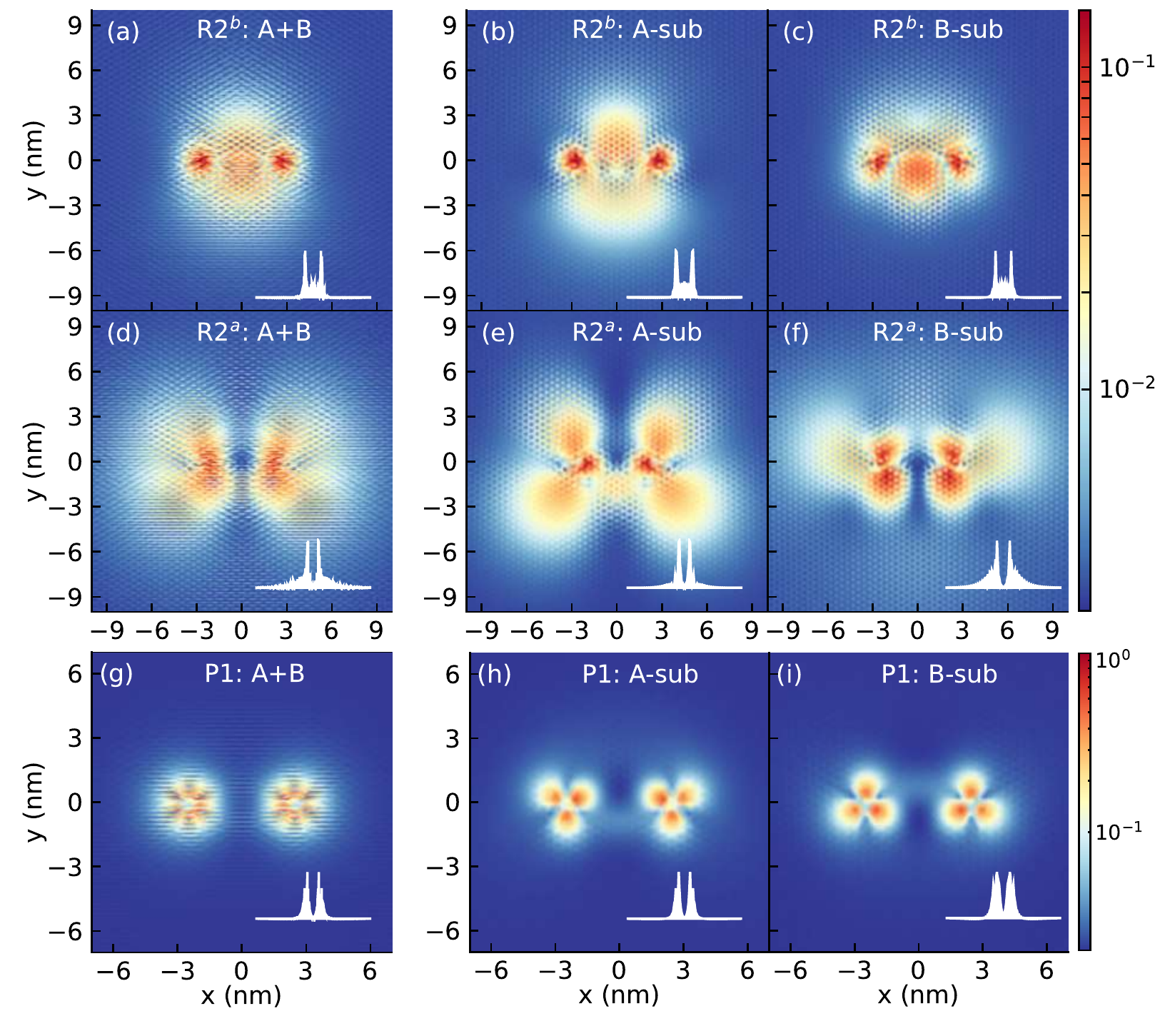}
        \caption{\label{fig:fig8} Spatial LDOS as in Fig.~\ref{fig:fig6} with A-A type vacancy with $R\approx 5$ nm for higher collapse states at
        $(E,\beta) = (-0.17~\text{eV},~1.5)$, $(-0.08~\text{eV},~1.5)$ and $(-0.3~\text{eV},~1.9)$ from bottom to top, respectively.}
    \end{figure}

\begin{figure}[htb!]
        \centering
        \includegraphics[width=\columnwidth]{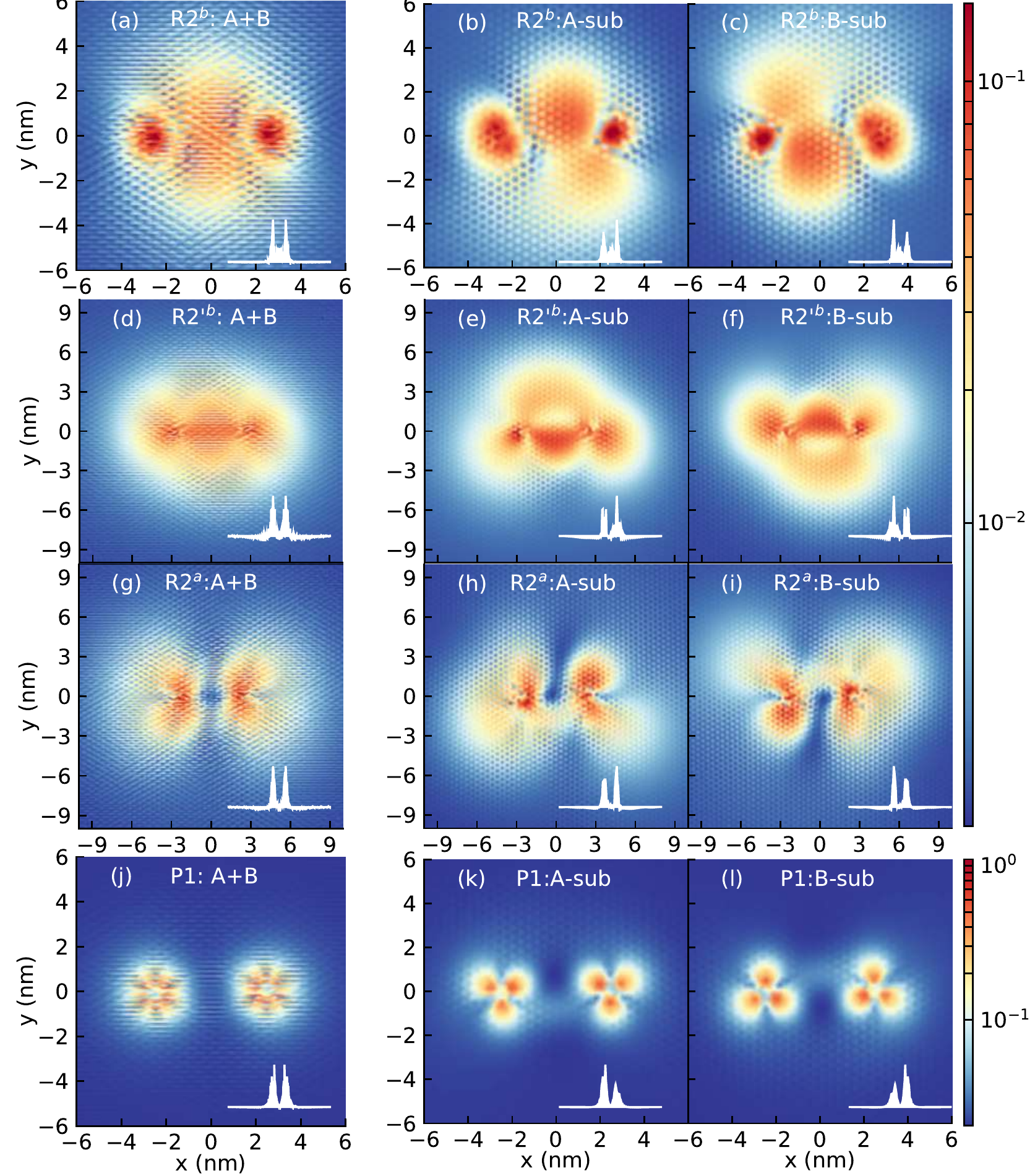}
        \caption{\label{fig:fig9} Spatial LDOS as in Fig.~\ref{fig:fig6}, for A-B type vacancy with $R\approx 5$ nm and higher collapse states at
        $(E,\beta) = (-0.17~\text{eV},~1.5)$, $(-0.13~\text{eV},~1.5)$, $(-0.08~\text{eV},~1.5)$ and $(-0.3~\text{eV},~1.9)$ from top to bottom, respectively.}
    \end{figure}

    For the purpose of further analysis, we plot in Figs.~\ref{fig:fig6}-\ref{fig:fig9} the spatial LDOS and the sublattice components of the different peaks labeled in Fig.~\ref{fig:fig2}. According to LCAO approximation~\cite{ref34, ref35}, the bonding and the anti-bonding molecular orbitals are respectively given by $\left|B_s\right\rangle=\sqrt{(1/2)} (\left|S_a\right\rangle + \left|S_b\right\rangle)$ and $\left|B_s\right\rangle=\sqrt{(1/2)} (\left|S_a\right\rangle - \left|S_b\right\rangle)$, with $\left|S_a\right\rangle$ and $\left|S_b\right\rangle$ the single atomic orbitals corresponding to the separate vacancies. The bonding (anti-bonding) state enhances (weakens) the intensity of the LDOS at the region between the two vacancies where the two wavefunctions overlap. From Fig.~\ref{fig:fig6}(a), Figs.~\ref{fig:fig7}(a, d), Fig.~\ref{fig:fig8}(a) and Figs.~\ref{fig:fig9}(a, d), the bonding feature is clearly observed, and can be distinguished from the anti-bonding one, shown in Figs.~\ref{fig:fig6}(d, g), Figs.~\ref{fig:fig7}(g, j), Fig.~\ref{fig:fig8}(d) and Fig.~\ref{fig:fig9}(g). 
    
    Separating individual sublattice components from the total spatial LDOS is useful for understanding the influence of the vacancies on the molecular collapse states. The spatial LDOS of the main molecular collapse resonances (R1$^b$, R1$^a$, R2$^b$ and R2$^a$) are similar to those found in the two charged impurity (placed above the graphene sheet) problem~\cite{ref20} and are therefore determined by the Coulomb potential. For the A-A type vacancy, the spatial LDOS distribution of R1$^b$ and R1$^a$ on A-sublattice in Figs.~\ref{fig:fig6}(b, e) resembles the B-sublattice components in Figs.~\ref{fig:fig6}(c, f), except around the vacancy where the LDOS is localized on the sublattice other than the one of the vacant site. The vacancy characteristic collapse peaks (R1'$^{a}$) is the vacancy introduced satellites of the peaks R1$^a$, and its spatial LDOS on the sublattice with the local majority of atoms (Fig.~\ref{fig:fig6}(i)) shows high density around the vacancy center spreading out along the zigzag directions. Meanwhile, the spatial LDOS on the other sublattice (Fig.~\ref{fig:fig6}(h)) shows similar distribution maps as the main corresponding molecular collapse states. Because A-A type defect with two removed A-atoms preserves the symmetry along the y axis, the spatial LDOS and the sublattice component distributions in Fig.~\ref{fig:fig6} and Fig.~\ref{fig:fig8} are symmetric about $x = 0$. If we compare the two defect types, two features become notable. Beside the R1'$^{b}$ and the R2'$^{b}$ peaks that can be distinguished in the case of a A-B type defect (see Fig.~\ref{fig:fig2}(b)), the two sublattice components of the spatial LDOS are now rotationally symmetric with respect to $180^\circ$ about $x = 0$ ($C_2$ symmetry), as can be seen in Figs.~\ref{fig:fig7} and~\ref{fig:fig9}. In addition, the spatial LDOS is helpful to identify the correspondence between the molecular collapse resonances and the well-known atomic orbitals from a Coulomb potential. The spatial LDOS in Figs.~\ref{fig:fig6}(a, d, g) and Figs.~\ref{fig:fig7}(a, d, g, j) displays high LDOS in the Coulomb centers and decreases along the radial direction, which is typical for the 1s atomic orbital. The spatial LDOS in Fig.~\ref{fig:fig8}(a) and Figs.~\ref{fig:fig9}(a, d) shows lower intensity rings outside the high LDOS center(if the two vacancies are separated by a larger distance, these rings, and the spatial distribution features of R1 states discussed above, are more clearly visible), which is analogous to the 2s atomic orbital. The anti-bonding molecular equivalent of the 2s atomic orbital in Fig.~\ref{fig:fig8}(d) and Fig.~\ref{fig:fig9}(g) shows two lower intensity islands instead of a ring-like feature. 

    All of the collapse states that we have discussed so far have angular momentum $m = 0$. To complement the study, in Fig.~\ref{fig:fig8}(g) and Fig.~\ref{fig:fig9}(j) we present the spatial LDOS and its sublattice components for the higher states with angular momentum $m = 1$. Notice that there are low LDOS intensity nodes at the position of the Coulomb centers which demonstrate that these states are the counterparts of the atomic p orbitals. The six-fold symmetry shape of the LDOS accounts for the honeycomb structure of graphene, i.e. it locks the lobes of the p-orbital along the crystallographic directions. To distinguish these states from other collapse peaks we plot the spatial LDOS in Fig.~\ref{fig:fig8}(g) and Fig.~\ref{fig:fig9}(j) for large values of effective charge $\beta$, thus overlap of the wavefuntions induced by the two impurities at the midpoint between the defects is reduced, and the bonding and the anti-bonding features are very weak.
    
    \begin{figure}[htb!]
        \centering
        \includegraphics[width=0.75\columnwidth]{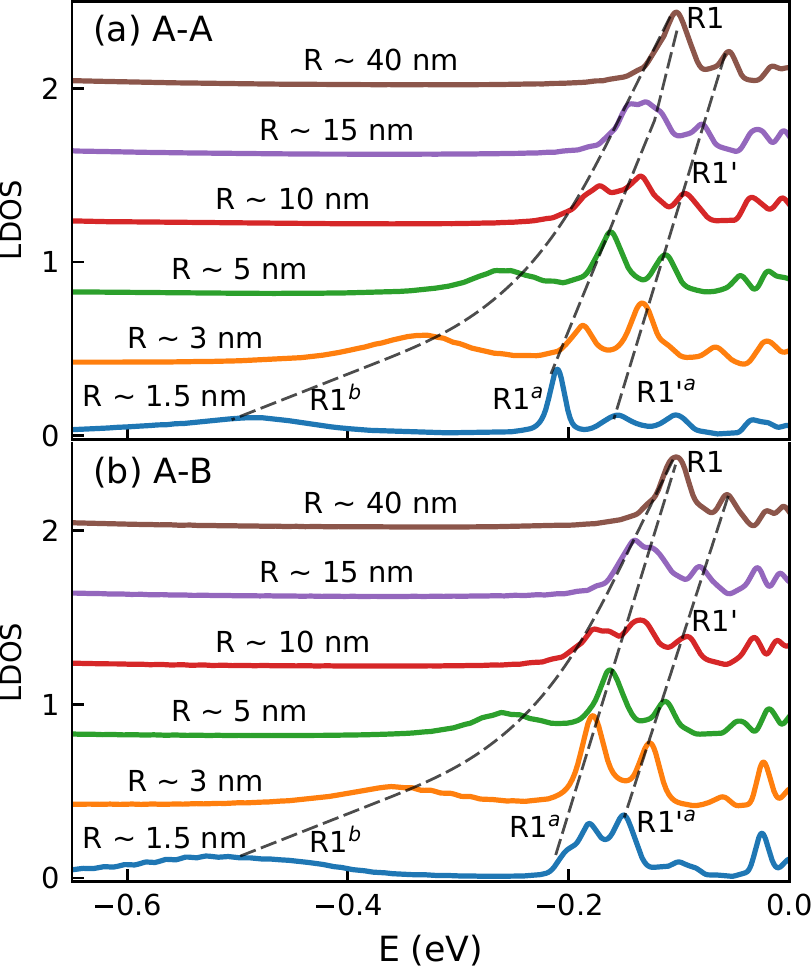}
        \caption{\label{fig:fig10} The LDOS as a function of energy for different values of inter-vacancy distance and $\beta = 1.1$, (a) A-A type vacancy, and (b) A-B type vacancy. Molecular bonding and anti-bonding collapse peaks merge at large distances into a single atomic collapse state. For this value of $\beta$ R1'$^{b}$ and R1$^{b}$ are already merged. Each line is offset with respect to the previous one for better distinction.}
    \end{figure}

    Next, we study the dependence of the molecular collapse resonances on the inter-charge distance. The LDOS as a function of energy for several values of $R$ is shown in Fig.~\ref{fig:fig10}, for effective charge $\beta = 1.1$. With increase of the vacancy separation, molecular bonding and anti-bonding collapse states come closer to each other, until they merge at high values of $R$ ($R\sim 15-40$ nm), when only the atomic collapse states are left. Because of a large effective charge $\beta$, both charges are individually supercritical, and the collapse states exist for all values of the separation distance $R$. 
    
\begin{figure}[htb!]
        \centering
        \includegraphics[width=\columnwidth]{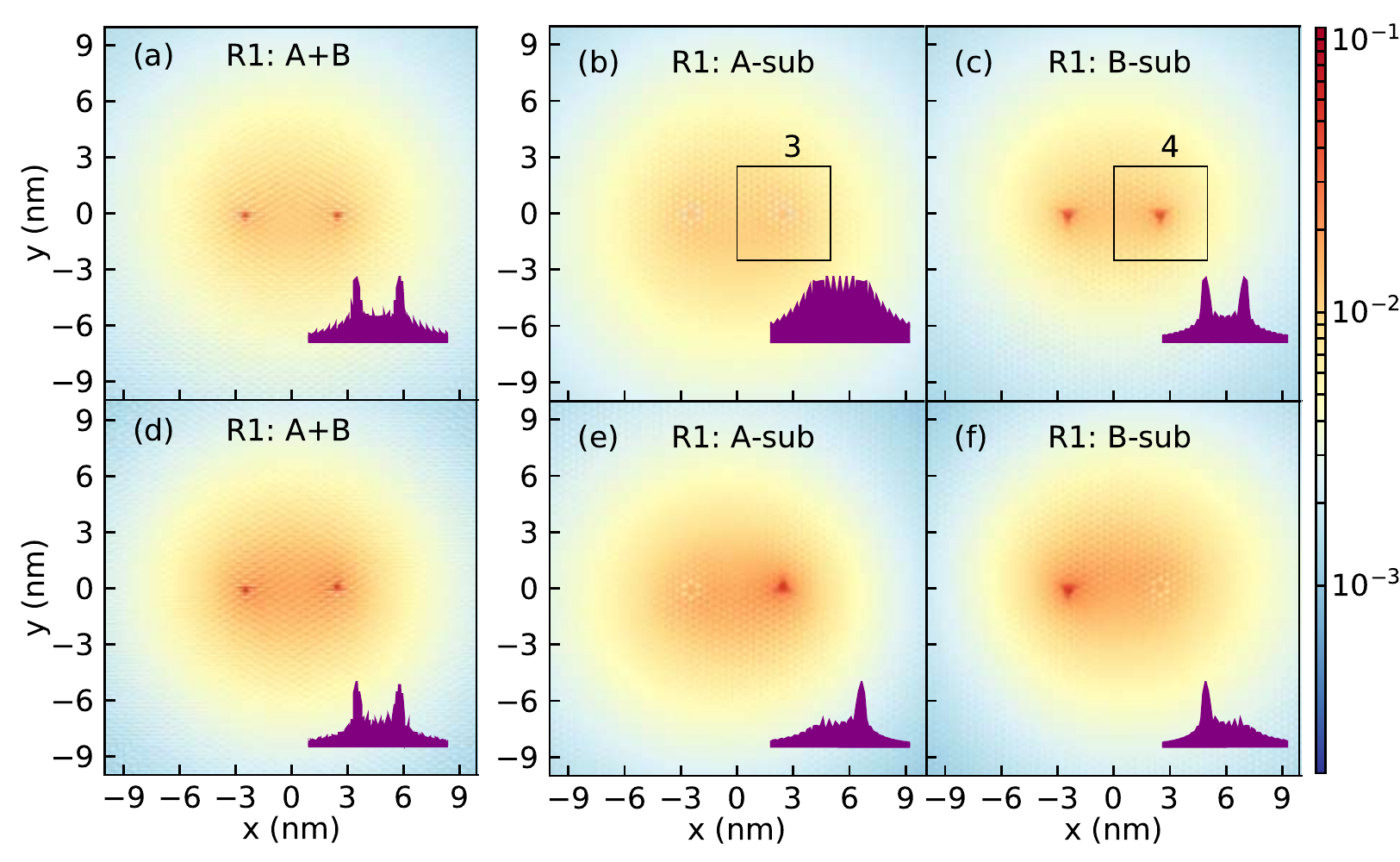}
        \caption{\label{fig:fig11} Spatial LDOS for the R1 resonance in frustrated atomic collapse regime. The total LDOS and the sublattice components, A and B, are shown in the left, middle and right panels, respectively. (a-c): A-A type vacancy, $E = -1.5$ meV, $\beta = 0.4$, (d-f): A-B type vacancy, $E=-9.3$ meV, $\beta = 0.47$, with $R\approx 5$ nm in both cases. }
    \end{figure}

\begin{figure}[htb!]
        \centering
        \includegraphics[width=0.75\columnwidth]{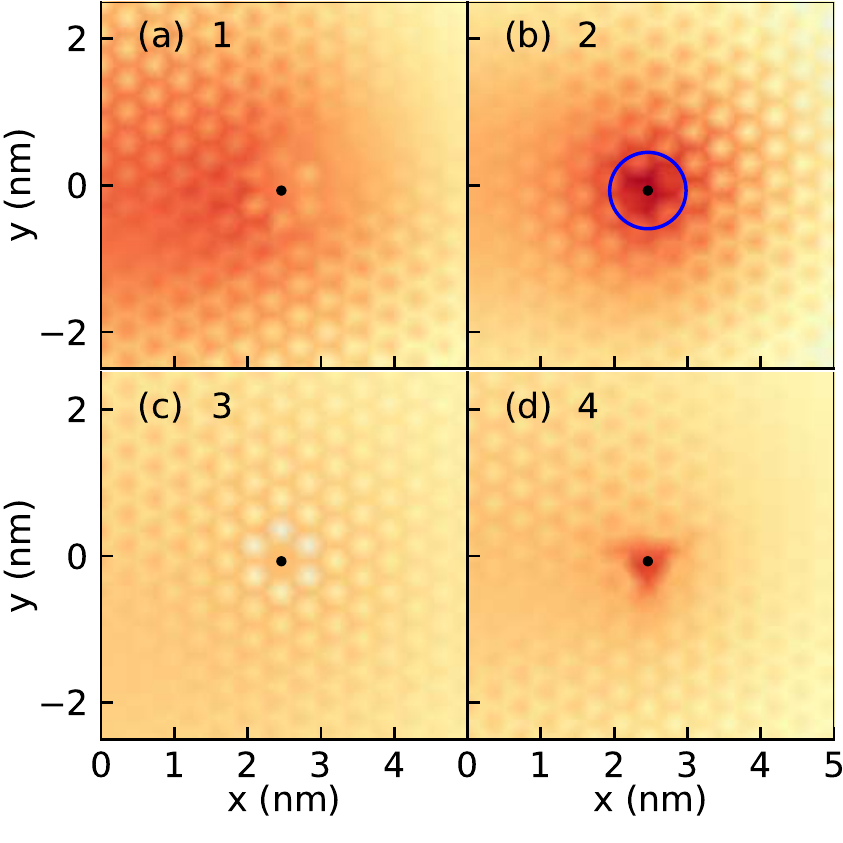}
        \caption{\label{fig:fig12} Spatial LDOS around a vacancy in the frustrated (bottom figures) and the molecular (top figures) collapse regimes. The areas are marked with black squares in Figs.~\ref{fig:fig6}(b, c) (top row) and Figs.~\ref{fig:fig11}(b, c) (bottom row).}
    \end{figure}

    When both charges are individually subcritical but together exceed the critical value, a phenomenon known as {\it frustrated atomic collapse}~\cite{ref20, ref25}, collapse states are present only for short inter-charge distances, and the spatial LDOS distribution of the molecular collapse resonances is different from the bonding or anti-bonding molecular orbitals. To further reflect on this difference, we plot in Fig.~\ref{fig:fig11} the spatial LDOS and the sublattice component of the R1 resonance in the frustrated collapse regime. The spatial LDOS shows two high intensity peaks centered around the vacancies, which mimics the formation of a bonding molecular orbital. In order to illustrate that these R1 states are indeed different from the bonding molecular orbital resonances, a comparison is presented in Fig.~\ref{fig:fig12}. The plot shows a comparison between the two regions marked with black boxes in Figs.~\ref{fig:fig6}(b, c) and Figs.~\ref{fig:fig11}(b, c), where the black dot represents the location of the vacancy. In Fig.~\ref{fig:fig12}(a), the A-sublattice component has an enhanced LDOS distribution on the side closer to the second vacancy, which differs from Fig.~\ref{fig:fig12}(c) where the A-sublattice LDOS is distributed almost uniformly. The purple inset in Fig.~\ref{fig:fig6}(b) displays two peaks around the vacancies with a slightly lower intensity on the outer side, while, the purple inset in Fig.~\ref{fig:fig11}(b) does not exhibit a clear reduced LDOS between the two vacancies. In Fig.~\ref{fig:fig12}(b), the B-sublattice component has a very high intensity inside the blue circle (a combined effect of Coulomb potential and vacancy induced localized state); outside the circle, the LDOS gradually decreases as one moves away from the vacancy (the dominant effect of the Coulomb potential). Considering again the LDOS along a line, the purple inset in Fig.~\ref{fig:fig6}(c) has a strongly suppressed LDOS between the two vacancies. In contrast, in Fig.~\ref{fig:fig12}(d), the states are highly concentrated in a triangular area, a localization solely caused by the presence of a vacancy. Outside the triangular area, the LDOS distribution is quasi-uniform. For A-B type vacancies, except for the different distribution over the sublattices, the conclusions are the same as for the A-A case. Furthermore, as the vacancy induced high LDOS centers only resemble a bonding molecular orbital, the spatial LDOS should reveal which collapse regime is at play, frustrated atomic collapse or the formation of the bonding molecular orbitals.

\section{Conclusion}\label{sec:concl}

    We studied how the sublattice symmetry manifests in the molecular collapse of a two charged vacancy system. By comparing the A-A with A-B vacancy types we showed that the global sublattice symmetry in the latter case results in distinguishable signatures with additional molecular collapse peaks. The main bonding and anti-bonding molecular collapse peaks of the two types are located at the same energy positions, while the energy of the vacancy induced satellite collapse peaks differs. As the individual charge, energy and separation distance increase, these additional molecular collapse peaks merge into the main features. 

    Depending on both the symmetry and the distance between the vacancies, the VP shows various properties. When the symmetry is broken a single peak appears. However, in case of A-B type system we find that for small $R$-values this VP is split into a bonding and anti-bonding feature. Because the splitting between the peaks depends on the overlap, these two peaks merge at higher values of the effective charge. If the distance is large, a single peak appears. We also found higher collapse states with angular momentum $m = 1$, they correspond to the atomic p orbitals, and exhibit a hexagonal symmetry of the spatial LDOS. 
    
    When both vacancies have each subcritical charge we identified a frustrated molecular collapse regime with decreasing distance between the two vacancies. In this regime the total system exhibits features similar to the molecular collapse state.
    
\begin{acknowledgments}
    This work was supported by the National Natural Science Foundation of China (Grant Nos. 61874038 and 61704040), National Key R\&D Program Grant 2018YFE0120000, the scholarship from China Scholarship Council (CSC: 201908330548), and TRANS2DTMD Flag-Era project.
\end{acknowledgments}

\bibliography{bibliography}

\begin{thebibliography}{34}%
\makeatletter
\providecommand \@ifxundefined [1]{%
 \@ifx{#1\undefined}
}%
\providecommand \@ifnum [1]{%
 \ifnum #1\expandafter \@firstoftwo
 \else \expandafter \@secondoftwo
 \fi
}%
\providecommand \@ifx [1]{%
 \ifx #1\expandafter \@firstoftwo
 \else \expandafter \@secondoftwo
 \fi
}%
\providecommand \natexlab [1]{#1}%
\providecommand \enquote  [1]{``#1''}%
\providecommand \bibnamefont  [1]{#1}%
\providecommand \bibfnamefont [1]{#1}%
\providecommand \citenamefont [1]{#1}%
\providecommand \href@noop [0]{\@secondoftwo}%
\providecommand \href [0]{\begingroup \@sanitize@url \@href}%
\providecommand \@href[1]{\@@startlink{#1}\@@href}%
\providecommand \@@href[1]{\endgroup#1\@@endlink}%
\providecommand \@sanitize@url [0]{\catcode `\\12\catcode `\$12\catcode
  `\&12\catcode `\#12\catcode `\^12\catcode `\_12\catcode `\%12\relax}%
\providecommand \@@startlink[1]{}%
\providecommand \@@endlink[0]{}%
\providecommand \url  [0]{\begingroup\@sanitize@url \@url }%
\providecommand \@url [1]{\endgroup\@href {#1}{\urlprefix }}%
\providecommand \urlprefix  [0]{URL }%
\providecommand \Eprint [0]{\href }%
\providecommand \doibase [0]{https://doi.org/}%
\providecommand \selectlanguage [0]{\@gobble}%
\providecommand \bibinfo  [0]{\@secondoftwo}%
\providecommand \bibfield  [0]{\@secondoftwo}%
\providecommand \translation [1]{[#1]}%
\providecommand \BibitemOpen [0]{}%
\providecommand \bibitemStop [0]{}%
\providecommand \bibitemNoStop [0]{.\EOS\space}%
\providecommand \EOS [0]{\spacefactor3000\relax}%
\providecommand \BibitemShut  [1]{\csname bibitem#1\endcsname}%
\let\auto@bib@innerbib\@empty
\bibitem [{\citenamefont {Greiner}\ \emph {et~al.}(1985)\citenamefont
  {Greiner}, \citenamefont {M\"{u}ller},\ and\ \citenamefont
  {Rafelski}}]{ref1}%
  \BibitemOpen
  \bibfield  {author} {\bibinfo {author} {\bibfnamefont {W.}~\bibnamefont
  {Greiner}}, \bibinfo {author} {\bibfnamefont {B.}~\bibnamefont
  {M\"{u}ller}},\ and\ \bibinfo {author} {\bibfnamefont {J.}~\bibnamefont
  {Rafelski}},\ }\href {https://doi.org/10.1007/978-3-642-82272-8} {\emph
  {\bibinfo {title} {Quantum Electrodynamics of Strong Fields}}}\ (\bibinfo
  {publisher} {Springer Berlin Heidelberg},\ \bibinfo {year}
  {1985})\BibitemShut {NoStop}%
\bibitem [{\citenamefont {Schweppe}\ \emph {et~al.}(1983)\citenamefont
  {Schweppe}, \citenamefont {Gruppe}, \citenamefont {Bethge}, \citenamefont
  {Bokemeyer}, \citenamefont {Cowan}, \citenamefont {Folger}, \citenamefont
  {Greenberg}, \citenamefont {Grein}, \citenamefont {Ito}, \citenamefont
  {Schule}, \citenamefont {Schwalm}, \citenamefont {Stiebing}, \citenamefont
  {Trautmann}, \citenamefont {Vincent},\ and\ \citenamefont
  {Waldschmidt}}]{ref2}%
  \BibitemOpen
  \bibfield  {author} {\bibinfo {author} {\bibfnamefont {J.}~\bibnamefont
  {Schweppe}}, \bibinfo {author} {\bibfnamefont {A.}~\bibnamefont {Gruppe}},
  \bibinfo {author} {\bibfnamefont {K.}~\bibnamefont {Bethge}}, \bibinfo
  {author} {\bibfnamefont {H.}~\bibnamefont {Bokemeyer}}, \bibinfo {author}
  {\bibfnamefont {T.}~\bibnamefont {Cowan}}, \bibinfo {author} {\bibfnamefont
  {H.}~\bibnamefont {Folger}}, \bibinfo {author} {\bibfnamefont {J.~S.}\
  \bibnamefont {Greenberg}}, \bibinfo {author} {\bibfnamefont {H.}~\bibnamefont
  {Grein}}, \bibinfo {author} {\bibfnamefont {S.}~\bibnamefont {Ito}}, \bibinfo
  {author} {\bibfnamefont {R.}~\bibnamefont {Schule}}, \bibinfo {author}
  {\bibfnamefont {D.}~\bibnamefont {Schwalm}}, \bibinfo {author} {\bibfnamefont
  {K.~E.}\ \bibnamefont {Stiebing}}, \bibinfo {author} {\bibfnamefont
  {N.}~\bibnamefont {Trautmann}}, \bibinfo {author} {\bibfnamefont
  {P.}~\bibnamefont {Vincent}},\ and\ \bibinfo {author} {\bibfnamefont
  {M.}~\bibnamefont {Waldschmidt}},\ }\bibfield  {title} {\bibinfo {title}
  {Observation of a peak structure in positron spectra from \textsc{U+C$_m$}
  collisions},\ }\href {https://doi.org/10.1103/PhysRevLett.51.2261} {\bibfield
   {journal} {\bibinfo  {journal} {Phys. Rev. Lett.}\ }\textbf {\bibinfo
  {volume} {51}},\ \bibinfo {pages} {2261} (\bibinfo {year}
  {1983})}\BibitemShut {NoStop}%
\bibitem [{\citenamefont {Cowan}\ \emph {et~al.}(1985)\citenamefont {Cowan},
  \citenamefont {Backe}, \citenamefont {Begemann}, \citenamefont {Bethge},
  \citenamefont {Bokemeyer}, \citenamefont {Folger}, \citenamefont {Greenberg},
  \citenamefont {Grein}, \citenamefont {Gruppe}, \citenamefont {Kido},
  \citenamefont {Kl\"uver}, \citenamefont {Schwalm}, \citenamefont {Schweppe},
  \citenamefont {Stiebing}, \citenamefont {Trautmann},\ and\ \citenamefont
  {Vincent}}]{ref3}%
  \BibitemOpen
  \bibfield  {author} {\bibinfo {author} {\bibfnamefont {T.}~\bibnamefont
  {Cowan}}, \bibinfo {author} {\bibfnamefont {H.}~\bibnamefont {Backe}},
  \bibinfo {author} {\bibfnamefont {M.}~\bibnamefont {Begemann}}, \bibinfo
  {author} {\bibfnamefont {K.}~\bibnamefont {Bethge}}, \bibinfo {author}
  {\bibfnamefont {H.}~\bibnamefont {Bokemeyer}}, \bibinfo {author}
  {\bibfnamefont {H.}~\bibnamefont {Folger}}, \bibinfo {author} {\bibfnamefont
  {J.~S.}\ \bibnamefont {Greenberg}}, \bibinfo {author} {\bibfnamefont
  {H.}~\bibnamefont {Grein}}, \bibinfo {author} {\bibfnamefont
  {A.}~\bibnamefont {Gruppe}}, \bibinfo {author} {\bibfnamefont
  {Y.}~\bibnamefont {Kido}}, \bibinfo {author} {\bibfnamefont {M.}~\bibnamefont
  {Kl\"uver}}, \bibinfo {author} {\bibfnamefont {D.}~\bibnamefont {Schwalm}},
  \bibinfo {author} {\bibfnamefont {J.}~\bibnamefont {Schweppe}}, \bibinfo
  {author} {\bibfnamefont {K.~E.}\ \bibnamefont {Stiebing}}, \bibinfo {author}
  {\bibfnamefont {N.}~\bibnamefont {Trautmann}},\ and\ \bibinfo {author}
  {\bibfnamefont {P.}~\bibnamefont {Vincent}},\ }\bibfield  {title} {\bibinfo
  {title} {Anomalous positron peaks from supercritical collision systems},\
  }\href {https://doi.org/10.1103/PhysRevLett.54.1761} {\bibfield  {journal}
  {\bibinfo  {journal} {Phys. Rev. Lett.}\ }\textbf {\bibinfo {volume} {54}},\
  \bibinfo {pages} {1761} (\bibinfo {year} {1985})}\BibitemShut {NoStop}%
\bibitem [{\citenamefont {Novoselov}\ \emph {et~al.}(2005)\citenamefont
  {Novoselov}, \citenamefont {Geim}, \citenamefont {Morozov}, \citenamefont
  {Jiang}, \citenamefont {Katsnelson}, \citenamefont {Grigorieva},
  \citenamefont {Dubonos},\ and\ \citenamefont {Firsov}}]{ref4}%
  \BibitemOpen
  \bibfield  {author} {\bibinfo {author} {\bibfnamefont {K.~S.}\ \bibnamefont
  {Novoselov}}, \bibinfo {author} {\bibfnamefont {A.~K.}\ \bibnamefont {Geim}},
  \bibinfo {author} {\bibfnamefont {S.~V.}\ \bibnamefont {Morozov}}, \bibinfo
  {author} {\bibfnamefont {D.}~\bibnamefont {Jiang}}, \bibinfo {author}
  {\bibfnamefont {M.~I.}\ \bibnamefont {Katsnelson}}, \bibinfo {author}
  {\bibfnamefont {I.~V.}\ \bibnamefont {Grigorieva}}, \bibinfo {author}
  {\bibfnamefont {S.~V.}\ \bibnamefont {Dubonos}},\ and\ \bibinfo {author}
  {\bibfnamefont {A.~A.}\ \bibnamefont {Firsov}},\ }\bibfield  {title}
  {\bibinfo {title} {Two-dimensional gas of massless dirac fermions in
  graphene},\ }\href {https://doi.org/10.1038/nature04233} {\bibfield
  {journal} {\bibinfo  {journal} {Nature}\ }\textbf {\bibinfo {volume} {438}},\
  \bibinfo {pages} {197} (\bibinfo {year} {2005})}\BibitemShut {NoStop}%
\bibitem [{\citenamefont {Khalilov}\ and\ \citenamefont {Ho}(1998)}]{ref5}%
  \BibitemOpen
  \bibfield  {author} {\bibinfo {author} {\bibfnamefont {V.~R.}\ \bibnamefont
  {Khalilov}}\ and\ \bibinfo {author} {\bibfnamefont {C.-L.}\ \bibnamefont
  {Ho}},\ }\bibfield  {title} {\bibinfo {title} {{Dirac} {electron} {in} a
  {Coulomb} {field} {in} (2$+$1) {dimensions}},\ }\href
  {https://doi.org/10.1142/s0217732398000668} {\bibfield  {journal} {\bibinfo
  {journal} {Modern Physics Letters A}\ }\textbf {\bibinfo {volume} {13}},\
  \bibinfo {pages} {615} (\bibinfo {year} {1998})}\BibitemShut {NoStop}%
\bibitem [{\citenamefont {Gorbar}\ \emph {et~al.}(2018)\citenamefont {Gorbar},
  \citenamefont {Gusynin},\ and\ \citenamefont {Sobol}}]{ref6}%
  \BibitemOpen
  \bibfield  {author} {\bibinfo {author} {\bibfnamefont {E.~V.}\ \bibnamefont
  {Gorbar}}, \bibinfo {author} {\bibfnamefont {V.~P.}\ \bibnamefont
  {Gusynin}},\ and\ \bibinfo {author} {\bibfnamefont {O.~O.}\ \bibnamefont
  {Sobol}},\ }\bibfield  {title} {\bibinfo {title} {Electron states in the
  field of charged impurities in two-dimensional dirac systems (review
  article)},\ }\href {https://doi.org/10.1063/1.5034149} {\bibfield  {journal}
  {\bibinfo  {journal} {Low Temp. Phys.}\ }\textbf {\bibinfo {volume} {44}},\
  \bibinfo {pages} {371} (\bibinfo {year} {2018})}\BibitemShut {NoStop}%
\bibitem [{\citenamefont {Zhang}\ and\ \citenamefont {Fogler}(2008)}]{ref7}%
  \BibitemOpen
  \bibfield  {author} {\bibinfo {author} {\bibfnamefont {L.~M.}\ \bibnamefont
  {Zhang}}\ and\ \bibinfo {author} {\bibfnamefont {M.~M.}\ \bibnamefont
  {Fogler}},\ }\bibfield  {title} {\bibinfo {title} {Nonlinear screening and
  ballistic transport in a graphene $p\mathrm{\text{\ensuremath{-}}}n$
  junction},\ }\href {https://doi.org/10.1103/PhysRevLett.100.116804}
  {\bibfield  {journal} {\bibinfo  {journal} {Phys. Rev. Lett.}\ }\textbf
  {\bibinfo {volume} {100}},\ \bibinfo {pages} {116804} (\bibinfo {year}
  {2008})}\BibitemShut {NoStop}%
\bibitem [{\citenamefont {Shytov}\ \emph
  {et~al.}(2007{\natexlab{a}})\citenamefont {Shytov}, \citenamefont
  {Katsnelson},\ and\ \citenamefont {Levitov}}]{ref8}%
  \BibitemOpen
  \bibfield  {author} {\bibinfo {author} {\bibfnamefont {A.~V.}\ \bibnamefont
  {Shytov}}, \bibinfo {author} {\bibfnamefont {M.~I.}\ \bibnamefont
  {Katsnelson}},\ and\ \bibinfo {author} {\bibfnamefont {L.~S.}\ \bibnamefont
  {Levitov}},\ }\bibfield  {title} {\bibinfo {title} {Atomic collapse and
  quasi--rydberg states in graphene},\ }\href
  {https://doi.org/10.1103/PhysRevLett.99.246802} {\bibfield  {journal}
  {\bibinfo  {journal} {Phys. Rev. Lett.}\ }\textbf {\bibinfo {volume} {99}},\
  \bibinfo {pages} {246802} (\bibinfo {year} {2007}{\natexlab{a}})}\BibitemShut
  {NoStop}%
\bibitem [{\citenamefont {Shytov}\ \emph
  {et~al.}(2007{\natexlab{b}})\citenamefont {Shytov}, \citenamefont
  {Katsnelson},\ and\ \citenamefont {Levitov}}]{ref9}%
  \BibitemOpen
  \bibfield  {author} {\bibinfo {author} {\bibfnamefont {A.~V.}\ \bibnamefont
  {Shytov}}, \bibinfo {author} {\bibfnamefont {M.~I.}\ \bibnamefont
  {Katsnelson}},\ and\ \bibinfo {author} {\bibfnamefont {L.~S.}\ \bibnamefont
  {Levitov}},\ }\bibfield  {title} {\bibinfo {title} {Vacuum polarization and
  screening of supercritical impurities in graphene},\ }\href
  {https://doi.org/10.1103/PhysRevLett.99.236801} {\bibfield  {journal}
  {\bibinfo  {journal} {Phys. Rev. Lett.}\ }\textbf {\bibinfo {volume} {99}},\
  \bibinfo {pages} {236801} (\bibinfo {year} {2007}{\natexlab{b}})}\BibitemShut
  {NoStop}%
\bibitem [{\citenamefont {Wang}\ \emph {et~al.}(2013)\citenamefont {Wang},
  \citenamefont {Wong}, \citenamefont {Shytov}, \citenamefont {Brar},
  \citenamefont {Choi}, \citenamefont {Wu}, \citenamefont {Tsai}, \citenamefont
  {Regan}, \citenamefont {Zettl}, \citenamefont {Kawakami}, \citenamefont
  {Louie}, \citenamefont {Levitov},\ and\ \citenamefont {Crommie}}]{ref10}%
  \BibitemOpen
  \bibfield  {author} {\bibinfo {author} {\bibfnamefont {Y.}~\bibnamefont
  {Wang}}, \bibinfo {author} {\bibfnamefont {D.}~\bibnamefont {Wong}}, \bibinfo
  {author} {\bibfnamefont {A.~V.}\ \bibnamefont {Shytov}}, \bibinfo {author}
  {\bibfnamefont {V.~W.}\ \bibnamefont {Brar}}, \bibinfo {author}
  {\bibfnamefont {S.}~\bibnamefont {Choi}}, \bibinfo {author} {\bibfnamefont
  {Q.}~\bibnamefont {Wu}}, \bibinfo {author} {\bibfnamefont {H.-Z.}\
  \bibnamefont {Tsai}}, \bibinfo {author} {\bibfnamefont {W.}~\bibnamefont
  {Regan}}, \bibinfo {author} {\bibfnamefont {A.}~\bibnamefont {Zettl}},
  \bibinfo {author} {\bibfnamefont {R.~K.}\ \bibnamefont {Kawakami}}, \bibinfo
  {author} {\bibfnamefont {S.~G.}\ \bibnamefont {Louie}}, \bibinfo {author}
  {\bibfnamefont {L.~S.}\ \bibnamefont {Levitov}},\ and\ \bibinfo {author}
  {\bibfnamefont {M.~F.}\ \bibnamefont {Crommie}},\ }\bibfield  {title}
  {\bibinfo {title} {Observing atomic collapse resonances in artificial nuclei
  on graphene},\ }\href {https://doi.org/10.1126/science.1234320} {\bibfield
  {journal} {\bibinfo  {journal} {Science}\ }\textbf {\bibinfo {volume}
  {340}},\ \bibinfo {pages} {734} (\bibinfo {year} {2013})}\BibitemShut
  {NoStop}%
\bibitem [{\citenamefont {Mao}\ \emph {et~al.}(2016)\citenamefont {Mao},
  \citenamefont {Jiang}, \citenamefont {Moldovan}, \citenamefont {Li},
  \citenamefont {Watanabe}, \citenamefont {Taniguchi}, \citenamefont {Masir},
  \citenamefont {Peeters},\ and\ \citenamefont {Andrei}}]{ref11}%
  \BibitemOpen
  \bibfield  {author} {\bibinfo {author} {\bibfnamefont {J.}~\bibnamefont
  {Mao}}, \bibinfo {author} {\bibfnamefont {Y.}~\bibnamefont {Jiang}}, \bibinfo
  {author} {\bibfnamefont {D.}~\bibnamefont {Moldovan}}, \bibinfo {author}
  {\bibfnamefont {G.}~\bibnamefont {Li}}, \bibinfo {author} {\bibfnamefont
  {K.}~\bibnamefont {Watanabe}}, \bibinfo {author} {\bibfnamefont
  {T.}~\bibnamefont {Taniguchi}}, \bibinfo {author} {\bibfnamefont {M.~R.}\
  \bibnamefont {Masir}}, \bibinfo {author} {\bibfnamefont {F.~M.}\ \bibnamefont
  {Peeters}},\ and\ \bibinfo {author} {\bibfnamefont {E.~Y.}\ \bibnamefont
  {Andrei}},\ }\bibfield  {title} {\bibinfo {title} {Realization of a tunable
  artificial atom at a supercritically charged vacancy in graphene},\ }\href
  {https://doi.org/10.1038/nphys3665} {\bibfield  {journal} {\bibinfo
  {journal} {Nat. Phys.}\ }\textbf {\bibinfo {volume} {12}},\ \bibinfo {pages}
  {545} (\bibinfo {year} {2016})}\BibitemShut {NoStop}%
\bibitem [{\citenamefont {Jiang}\ \emph {et~al.}(2017)\citenamefont {Jiang},
  \citenamefont {Mao}, \citenamefont {Moldovan}, \citenamefont {Masir},
  \citenamefont {Li}, \citenamefont {Watanabe}, \citenamefont {Taniguchi},
  \citenamefont {Peeters},\ and\ \citenamefont {Andrei}}]{ref12}%
  \BibitemOpen
  \bibfield  {author} {\bibinfo {author} {\bibfnamefont {Y.}~\bibnamefont
  {Jiang}}, \bibinfo {author} {\bibfnamefont {J.}~\bibnamefont {Mao}}, \bibinfo
  {author} {\bibfnamefont {D.}~\bibnamefont {Moldovan}}, \bibinfo {author}
  {\bibfnamefont {M.~R.}\ \bibnamefont {Masir}}, \bibinfo {author}
  {\bibfnamefont {G.}~\bibnamefont {Li}}, \bibinfo {author} {\bibfnamefont
  {K.}~\bibnamefont {Watanabe}}, \bibinfo {author} {\bibfnamefont
  {T.}~\bibnamefont {Taniguchi}}, \bibinfo {author} {\bibfnamefont {F.~M.}\
  \bibnamefont {Peeters}},\ and\ \bibinfo {author} {\bibfnamefont {E.~Y.}\
  \bibnamefont {Andrei}},\ }\bibfield  {title} {\bibinfo {title} {Tuning a
  circular~p{\textendash}n~junction in graphene from quantum confinement to
  optical guiding},\ }\href {https://doi.org/10.1038/nnano.2017.181} {\bibfield
   {journal} {\bibinfo  {journal} {Nat. Nanotechnol.}\ }\textbf {\bibinfo
  {volume} {12}},\ \bibinfo {pages} {1045} (\bibinfo {year}
  {2017})}\BibitemShut {NoStop}%
\bibitem [{\citenamefont {Fogler}\ \emph {et~al.}(2007)\citenamefont {Fogler},
  \citenamefont {Novikov},\ and\ \citenamefont {Shklovskii}}]{ref13}%
  \BibitemOpen
  \bibfield  {author} {\bibinfo {author} {\bibfnamefont {M.~M.}\ \bibnamefont
  {Fogler}}, \bibinfo {author} {\bibfnamefont {D.~S.}\ \bibnamefont
  {Novikov}},\ and\ \bibinfo {author} {\bibfnamefont {B.~I.}\ \bibnamefont
  {Shklovskii}},\ }\bibfield  {title} {\bibinfo {title} {Screening of a
  hypercritical charge in graphene},\ }\href
  {https://doi.org/10.1103/PhysRevB.76.233402} {\bibfield  {journal} {\bibinfo
  {journal} {Phys. Rev. B}\ }\textbf {\bibinfo {volume} {76}},\ \bibinfo
  {pages} {233402} (\bibinfo {year} {2007})}\BibitemShut {NoStop}%
\bibitem [{\citenamefont {Terekhov}\ \emph {et~al.}(2008)\citenamefont
  {Terekhov}, \citenamefont {Milstein}, \citenamefont {Kotov},\ and\
  \citenamefont {Sushkov}}]{ref14}%
  \BibitemOpen
  \bibfield  {author} {\bibinfo {author} {\bibfnamefont {I.~S.}\ \bibnamefont
  {Terekhov}}, \bibinfo {author} {\bibfnamefont {A.~I.}\ \bibnamefont
  {Milstein}}, \bibinfo {author} {\bibfnamefont {V.~N.}\ \bibnamefont
  {Kotov}},\ and\ \bibinfo {author} {\bibfnamefont {O.~P.}\ \bibnamefont
  {Sushkov}},\ }\bibfield  {title} {\bibinfo {title} {Screening of coulomb
  impurities in graphene},\ }\href
  {https://doi.org/10.1103/PhysRevLett.100.076803} {\bibfield  {journal}
  {\bibinfo  {journal} {Phys. Rev. Lett.}\ }\textbf {\bibinfo {volume} {100}},\
  \bibinfo {pages} {076803} (\bibinfo {year} {2008})}\BibitemShut {NoStop}%
\bibitem [{\citenamefont {Pereira}\ \emph {et~al.}(2007)\citenamefont
  {Pereira}, \citenamefont {Nilsson},\ and\ \citenamefont
  {Castro~Neto}}]{ref15}%
  \BibitemOpen
  \bibfield  {author} {\bibinfo {author} {\bibfnamefont {V.~M.}\ \bibnamefont
  {Pereira}}, \bibinfo {author} {\bibfnamefont {J.}~\bibnamefont {Nilsson}},\
  and\ \bibinfo {author} {\bibfnamefont {A.~H.}\ \bibnamefont {Castro~Neto}},\
  }\bibfield  {title} {\bibinfo {title} {Coulomb impurity problem in
  graphene},\ }\href {https://doi.org/10.1103/PhysRevLett.99.166802} {\bibfield
   {journal} {\bibinfo  {journal} {Phys. Rev. Lett.}\ }\textbf {\bibinfo
  {volume} {99}},\ \bibinfo {pages} {166802} (\bibinfo {year}
  {2007})}\BibitemShut {NoStop}%
\bibitem [{\citenamefont {Neto}\ \emph {et~al.}(2009)\citenamefont {Neto},
  \citenamefont {Kotov}, \citenamefont {Nilsson}, \citenamefont {Pereira},
  \citenamefont {Peres},\ and\ \citenamefont {Uchoa}}]{ref16}%
  \BibitemOpen
  \bibfield  {author} {\bibinfo {author} {\bibfnamefont {A.~C.}\ \bibnamefont
  {Neto}}, \bibinfo {author} {\bibfnamefont {V.}~\bibnamefont {Kotov}},
  \bibinfo {author} {\bibfnamefont {J.}~\bibnamefont {Nilsson}}, \bibinfo
  {author} {\bibfnamefont {V.}~\bibnamefont {Pereira}}, \bibinfo {author}
  {\bibfnamefont {N.}~\bibnamefont {Peres}},\ and\ \bibinfo {author}
  {\bibfnamefont {B.}~\bibnamefont {Uchoa}},\ }\bibfield  {title} {\bibinfo
  {title} {Adatoms in graphene},\ }\href
  {https://doi.org/10.1016/j.ssc.2009.02.040} {\bibfield  {journal} {\bibinfo
  {journal} {Solid State Commun.}\ }\textbf {\bibinfo {volume} {149}},\
  \bibinfo {pages} {1094} (\bibinfo {year} {2009})}\BibitemShut {NoStop}%
\bibitem [{\citenamefont {Novikov}(2007)}]{ref17}%
  \BibitemOpen
  \bibfield  {author} {\bibinfo {author} {\bibfnamefont {D.~S.}\ \bibnamefont
  {Novikov}},\ }\bibfield  {title} {\bibinfo {title} {Elastic scattering theory
  and transport in graphene},\ }\href
  {https://doi.org/10.1103/PhysRevB.76.245435} {\bibfield  {journal} {\bibinfo
  {journal} {Phys. Rev. B}\ }\textbf {\bibinfo {volume} {76}},\ \bibinfo
  {pages} {245435} (\bibinfo {year} {2007})}\BibitemShut {NoStop}%
\bibitem [{\citenamefont {Kotov}\ \emph {et~al.}(2012)\citenamefont {Kotov},
  \citenamefont {Uchoa}, \citenamefont {Pereira}, \citenamefont {Guinea},\ and\
  \citenamefont {Castro~Neto}}]{ref18}%
  \BibitemOpen
  \bibfield  {author} {\bibinfo {author} {\bibfnamefont {V.~N.}\ \bibnamefont
  {Kotov}}, \bibinfo {author} {\bibfnamefont {B.}~\bibnamefont {Uchoa}},
  \bibinfo {author} {\bibfnamefont {V.~M.}\ \bibnamefont {Pereira}}, \bibinfo
  {author} {\bibfnamefont {F.}~\bibnamefont {Guinea}},\ and\ \bibinfo {author}
  {\bibfnamefont {A.~H.}\ \bibnamefont {Castro~Neto}},\ }\bibfield  {title}
  {\bibinfo {title} {Electron-electron interactions in graphene: Current status
  and perspectives},\ }\href {https://doi.org/10.1103/RevModPhys.84.1067}
  {\bibfield  {journal} {\bibinfo  {journal} {Rev. Mod. Phys.}\ }\textbf
  {\bibinfo {volume} {84}},\ \bibinfo {pages} {1067} (\bibinfo {year}
  {2012})}\BibitemShut {NoStop}%
\bibitem [{\citenamefont {Moldovan}\ \emph
  {et~al.}(2017{\natexlab{a}})\citenamefont {Moldovan}, \citenamefont {Masir},\
  and\ \citenamefont {Peeters}}]{ref19}%
  \BibitemOpen
  \bibfield  {author} {\bibinfo {author} {\bibfnamefont {D.}~\bibnamefont
  {Moldovan}}, \bibinfo {author} {\bibfnamefont {M.~R.}\ \bibnamefont
  {Masir}},\ and\ \bibinfo {author} {\bibfnamefont {F.~M.}\ \bibnamefont
  {Peeters}},\ }\bibfield  {title} {\bibinfo {title} {Magnetic field dependence
  of the atomic collapse state in graphene},\ }\href
  {https://doi.org/10.1088/2053-1583/aa9647} {\bibfield  {journal} {\bibinfo
  {journal} {2D Mater.}\ }\textbf {\bibinfo {volume} {5}},\ \bibinfo {pages}
  {015017} (\bibinfo {year} {2017}{\natexlab{a}})}\BibitemShut {NoStop}%
\bibitem [{\citenamefont {Pottelberge}\ \emph {et~al.}(2019)\citenamefont
  {Pottelberge}, \citenamefont {Moldovan}, \citenamefont {Milovanovi{\'{c}}},\
  and\ \citenamefont {Peeters}}]{ref20}%
  \BibitemOpen
  \bibfield  {author} {\bibinfo {author} {\bibfnamefont {R.~V.}\ \bibnamefont
  {Pottelberge}}, \bibinfo {author} {\bibfnamefont {D.}~\bibnamefont
  {Moldovan}}, \bibinfo {author} {\bibfnamefont {S.~P.}\ \bibnamefont
  {Milovanovi{\'{c}}}},\ and\ \bibinfo {author} {\bibfnamefont {F.~M.}\
  \bibnamefont {Peeters}},\ }\bibfield  {title} {\bibinfo {title} {Molecular
  collapse in monolayer graphene},\ }\href
  {https://doi.org/10.1088/2053-1583/ab3feb} {\bibfield  {journal} {\bibinfo
  {journal} {2D Mater.}\ }\textbf {\bibinfo {volume} {6}},\ \bibinfo {pages}
  {045047} (\bibinfo {year} {2019})}\BibitemShut {NoStop}%
\bibitem [{\citenamefont {De~Martino}\ \emph {et~al.}(2014)\citenamefont
  {De~Martino}, \citenamefont {Kl\"opfer}, \citenamefont {Matrasulov},\ and\
  \citenamefont {Egger}}]{ref21}%
  \BibitemOpen
  \bibfield  {author} {\bibinfo {author} {\bibfnamefont {A.}~\bibnamefont
  {De~Martino}}, \bibinfo {author} {\bibfnamefont {D.}~\bibnamefont
  {Kl\"opfer}}, \bibinfo {author} {\bibfnamefont {D.}~\bibnamefont
  {Matrasulov}},\ and\ \bibinfo {author} {\bibfnamefont {R.}~\bibnamefont
  {Egger}},\ }\bibfield  {title} {\bibinfo {title} {Electric-dipole-induced
  universality for dirac fermions in graphene},\ }\href
  {https://doi.org/10.1103/PhysRevLett.112.186603} {\bibfield  {journal}
  {\bibinfo  {journal} {Phys. Rev. Lett.}\ }\textbf {\bibinfo {volume} {112}},\
  \bibinfo {pages} {186603} (\bibinfo {year} {2014})}\BibitemShut {NoStop}%
\bibitem [{\citenamefont {Kl\"{o}pfer}\ \emph {et~al.}(2014)\citenamefont
  {Kl\"{o}pfer}, \citenamefont {Martino}, \citenamefont {Matrasulov},\ and\
  \citenamefont {Egger}}]{ref22}%
  \BibitemOpen
  \bibfield  {author} {\bibinfo {author} {\bibfnamefont {D.}~\bibnamefont
  {Kl\"{o}pfer}}, \bibinfo {author} {\bibfnamefont {A.~D.}\ \bibnamefont
  {Martino}}, \bibinfo {author} {\bibfnamefont {D.~U.}\ \bibnamefont
  {Matrasulov}},\ and\ \bibinfo {author} {\bibfnamefont {R.}~\bibnamefont
  {Egger}},\ }\bibfield  {title} {\bibinfo {title} {Scattering theory and
  ground-state energy of \textsc{D}irac fermions in graphene with two
  \textsc{C}oulomb impurities},\ }\href
  {https://doi.org/10.1140/epjb/e2014-50414-8} {\bibfield  {journal} {\bibinfo
  {journal} {Eur. Phys. J. B}\ }\textbf {\bibinfo {volume} {87}},\ \bibinfo
  {pages} {187} (\bibinfo {year} {2014})}\BibitemShut {NoStop}%
\bibitem [{\citenamefont {Gorbar}\ \emph {et~al.}(2015)\citenamefont {Gorbar},
  \citenamefont {Gusynin},\ and\ \citenamefont {Sobol}}]{ref23}%
  \BibitemOpen
  \bibfield  {author} {\bibinfo {author} {\bibfnamefont {E.~V.}\ \bibnamefont
  {Gorbar}}, \bibinfo {author} {\bibfnamefont {V.~P.}\ \bibnamefont
  {Gusynin}},\ and\ \bibinfo {author} {\bibfnamefont {O.~O.}\ \bibnamefont
  {Sobol}},\ }\bibfield  {title} {\bibinfo {title} {Supercriticality of novel
  type induced by electric dipole in gapped graphene},\ }\href
  {https://doi.org/10.1103/PhysRevB.92.235417} {\bibfield  {journal} {\bibinfo
  {journal} {Phys. Rev. B}\ }\textbf {\bibinfo {volume} {92}},\ \bibinfo
  {pages} {235417} (\bibinfo {year} {2015})}\BibitemShut {NoStop}%
\bibitem [{\citenamefont {Van~Pottelberge}\ \emph {et~al.}(2018)\citenamefont
  {Van~Pottelberge}, \citenamefont {Van~Duppen},\ and\ \citenamefont
  {Peeters}}]{ref24}%
  \BibitemOpen
  \bibfield  {author} {\bibinfo {author} {\bibfnamefont {R.}~\bibnamefont
  {Van~Pottelberge}}, \bibinfo {author} {\bibfnamefont {B.}~\bibnamefont
  {Van~Duppen}},\ and\ \bibinfo {author} {\bibfnamefont {F.~M.}\ \bibnamefont
  {Peeters}},\ }\bibfield  {title} {\bibinfo {title} {Electrical dipole on
  gapped graphene: Bound states and atomic collapse},\ }\href
  {https://doi.org/10.1103/PhysRevB.98.165420} {\bibfield  {journal} {\bibinfo
  {journal} {Phys. Rev. B}\ }\textbf {\bibinfo {volume} {98}},\ \bibinfo
  {pages} {165420} (\bibinfo {year} {2018})}\BibitemShut {NoStop}%
\bibitem [{\citenamefont {Lu}\ \emph {et~al.}(2019)\citenamefont {Lu},
  \citenamefont {Tsai}, \citenamefont {Tatan}, \citenamefont {Wickenburg},
  \citenamefont {Omrani}, \citenamefont {Wong}, \citenamefont {Riss},
  \citenamefont {Piatti}, \citenamefont {Watanabe}, \citenamefont {Taniguchi},
  \citenamefont {Zettl}, \citenamefont {Pereira},\ and\ \citenamefont
  {Crommie}}]{ref25}%
  \BibitemOpen
  \bibfield  {author} {\bibinfo {author} {\bibfnamefont {J.}~\bibnamefont
  {Lu}}, \bibinfo {author} {\bibfnamefont {H.-Z.}\ \bibnamefont {Tsai}},
  \bibinfo {author} {\bibfnamefont {A.~N.}\ \bibnamefont {Tatan}}, \bibinfo
  {author} {\bibfnamefont {S.}~\bibnamefont {Wickenburg}}, \bibinfo {author}
  {\bibfnamefont {A.~A.}\ \bibnamefont {Omrani}}, \bibinfo {author}
  {\bibfnamefont {D.}~\bibnamefont {Wong}}, \bibinfo {author} {\bibfnamefont
  {A.}~\bibnamefont {Riss}}, \bibinfo {author} {\bibfnamefont {E.}~\bibnamefont
  {Piatti}}, \bibinfo {author} {\bibfnamefont {K.}~\bibnamefont {Watanabe}},
  \bibinfo {author} {\bibfnamefont {T.}~\bibnamefont {Taniguchi}}, \bibinfo
  {author} {\bibfnamefont {A.}~\bibnamefont {Zettl}}, \bibinfo {author}
  {\bibfnamefont {V.~M.}\ \bibnamefont {Pereira}},\ and\ \bibinfo {author}
  {\bibfnamefont {M.~F.}\ \bibnamefont {Crommie}},\ }\bibfield  {title}
  {\bibinfo {title} {Frustrated supercritical collapse in tunable charge arrays
  on graphene},\ }\href {https://doi.org/10.1038/s41467-019-08371-2} {\bibfield
   {journal} {\bibinfo  {journal} {Nat. Commun.}\ }\textbf {\bibinfo {volume}
  {10}},\ \bibinfo {pages} {477} (\bibinfo {year} {2019})}\BibitemShut
  {NoStop}%
\bibitem [{\citenamefont {Liu}\ \emph {et~al.}(2014)\citenamefont {Liu},
  \citenamefont {Weinert},\ and\ \citenamefont {Li}}]{ref26}%
  \BibitemOpen
  \bibfield  {author} {\bibinfo {author} {\bibfnamefont {Y.}~\bibnamefont
  {Liu}}, \bibinfo {author} {\bibfnamefont {M.}~\bibnamefont {Weinert}},\ and\
  \bibinfo {author} {\bibfnamefont {L.}~\bibnamefont {Li}},\ }\bibfield
  {title} {\bibinfo {title} {Determining charge state of graphene vacancy by
  noncontact atomic force microscopy and first-principles calculations},\
  }\href {https://doi.org/10.1088/0957-4484/26/3/035702} {\bibfield  {journal}
  {\bibinfo  {journal} {Nanotechnology}\ }\textbf {\bibinfo {volume} {26}},\
  \bibinfo {pages} {035702} (\bibinfo {year} {2014})}\BibitemShut {NoStop}%
\bibitem [{\citenamefont {Moldovan}\ and\ \citenamefont
  {Peeters}(2016)}]{ref27}%
  \BibitemOpen
  \bibfield  {author} {\bibinfo {author} {\bibfnamefont {D.}~\bibnamefont
  {Moldovan}}\ and\ \bibinfo {author} {\bibfnamefont {F.~M.}\ \bibnamefont
  {Peeters}},\ }\bibfield  {title} {\bibinfo {title} {Atomic collapse in
  graphene},\ }in\ \href {https://doi.org/10.1007/978-94-017-7593-9_1} {\emph
  {\bibinfo {booktitle} {Nanomaterials for Security}}}\ (\bibinfo  {publisher}
  {Springer Netherlands},\ \bibinfo {year} {2016})\ pp.\ \bibinfo {pages}
  {3--17}\BibitemShut {NoStop}%
\bibitem [{\citenamefont {Lehtinen}\ \emph {et~al.}(2010)\citenamefont
  {Lehtinen}, \citenamefont {Kotakoski}, \citenamefont {Krasheninnikov},
  \citenamefont {Tolvanen}, \citenamefont {Nordlund},\ and\ \citenamefont
  {Keinonen}}]{ref28}%
  \BibitemOpen
  \bibfield  {author} {\bibinfo {author} {\bibfnamefont {O.}~\bibnamefont
  {Lehtinen}}, \bibinfo {author} {\bibfnamefont {J.}~\bibnamefont {Kotakoski}},
  \bibinfo {author} {\bibfnamefont {A.~V.}\ \bibnamefont {Krasheninnikov}},
  \bibinfo {author} {\bibfnamefont {A.}~\bibnamefont {Tolvanen}}, \bibinfo
  {author} {\bibfnamefont {K.}~\bibnamefont {Nordlund}},\ and\ \bibinfo
  {author} {\bibfnamefont {J.}~\bibnamefont {Keinonen}},\ }\bibfield  {title}
  {\bibinfo {title} {Effects of ion bombardment on a two-dimensional target:
  Atomistic simulations of graphene irradiation},\ }\href
  {https://doi.org/10.1103/PhysRevB.81.153401} {\bibfield  {journal} {\bibinfo
  {journal} {Phys. Rev. B}\ }\textbf {\bibinfo {volume} {81}},\ \bibinfo
  {pages} {153401} (\bibinfo {year} {2010})}\BibitemShut {NoStop}%
\bibitem [{\citenamefont {Pereira}\ \emph {et~al.}(2006)\citenamefont
  {Pereira}, \citenamefont {Guinea}, \citenamefont {Lopes~dos Santos},
  \citenamefont {Peres},\ and\ \citenamefont {Castro~Neto}}]{ref29}%
  \BibitemOpen
  \bibfield  {author} {\bibinfo {author} {\bibfnamefont {V.~M.}\ \bibnamefont
  {Pereira}}, \bibinfo {author} {\bibfnamefont {F.}~\bibnamefont {Guinea}},
  \bibinfo {author} {\bibfnamefont {J.~M.~B.}\ \bibnamefont {Lopes~dos
  Santos}}, \bibinfo {author} {\bibfnamefont {N.~M.~R.}\ \bibnamefont
  {Peres}},\ and\ \bibinfo {author} {\bibfnamefont {A.~H.}\ \bibnamefont
  {Castro~Neto}},\ }\bibfield  {title} {\bibinfo {title} {Disorder induced
  localized states in graphene},\ }\href
  {https://doi.org/10.1103/PhysRevLett.96.036801} {\bibfield  {journal}
  {\bibinfo  {journal} {Phys. Rev. Lett.}\ }\textbf {\bibinfo {volume} {96}},\
  \bibinfo {pages} {036801} (\bibinfo {year} {2006})}\BibitemShut {NoStop}%
\bibitem [{\citenamefont {Pereira}\ \emph {et~al.}(2008)\citenamefont
  {Pereira}, \citenamefont {Lopes~dos Santos},\ and\ \citenamefont
  {Castro~Neto}}]{ref30}%
  \BibitemOpen
  \bibfield  {author} {\bibinfo {author} {\bibfnamefont {V.~M.}\ \bibnamefont
  {Pereira}}, \bibinfo {author} {\bibfnamefont {J.~M.~B.}\ \bibnamefont
  {Lopes~dos Santos}},\ and\ \bibinfo {author} {\bibfnamefont {A.~H.}\
  \bibnamefont {Castro~Neto}},\ }\bibfield  {title} {\bibinfo {title} {Modeling
  disorder in graphene},\ }\href {https://doi.org/10.1103/PhysRevB.77.115109}
  {\bibfield  {journal} {\bibinfo  {journal} {Phys. Rev. B}\ }\textbf {\bibinfo
  {volume} {77}},\ \bibinfo {pages} {115109} (\bibinfo {year}
  {2008})}\BibitemShut {NoStop}%
\bibitem [{\citenamefont {Moldovan}\ \emph
  {et~al.}(2017{\natexlab{b}})\citenamefont {Moldovan}, \citenamefont
  {Anđelković},\ and\ \citenamefont {Peeters}}]{ref31}%
  \BibitemOpen
  \bibfield  {author} {\bibinfo {author} {\bibfnamefont {D.}~\bibnamefont
  {Moldovan}}, \bibinfo {author} {\bibfnamefont {M.}~\bibnamefont
  {Anđelković}},\ and\ \bibinfo {author} {\bibfnamefont {F.~M.}\ \bibnamefont
  {Peeters}},\ }\href {https://doi.org/10.5281/zenodo.826942} {\bibinfo {title}
  {Pybinding v0.9.4: A python package for tight-binding calculations}}
  (\bibinfo {year} {2017}{\natexlab{b}})\BibitemShut {NoStop}%
\bibitem [{\citenamefont {Zeldovich}\ and\ \citenamefont
  {Popov}(1972)}]{ref32}%
  \BibitemOpen
  \bibfield  {author} {\bibinfo {author} {\bibfnamefont {Y.~B.}\ \bibnamefont
  {Zeldovich}}\ and\ \bibinfo {author} {\bibfnamefont {V.~S.}\ \bibnamefont
  {Popov}},\ }\bibfield  {title} {\bibinfo {title} {{Electronic} {structure}
  {of} {superheavy} {atoms}},\ }\href
  {https://doi.org/10.1070/pu1972v014n06abeh004735} {\bibfield  {journal}
  {\bibinfo  {journal} {Sov. Phys. Uspekhi}\ }\textbf {\bibinfo {volume}
  {14}},\ \bibinfo {pages} {673} (\bibinfo {year} {1972})}\BibitemShut
  {NoStop}%
\bibitem [{\citenamefont {Lennard-Jones}(1929)}]{ref34}%
  \BibitemOpen
  \bibfield  {author} {\bibinfo {author} {\bibfnamefont {J.~E.}\ \bibnamefont
  {Lennard-Jones}},\ }\bibfield  {title} {\bibinfo {title} {The electronic
  structure of some diatomic molecules},\ }\href
  {https://doi.org/10.1039/tf9292500668} {\bibfield  {journal} {\bibinfo
  {journal} {Transactions of the Faraday Society}\ }\textbf {\bibinfo {volume}
  {25}},\ \bibinfo {pages} {668} (\bibinfo {year} {1929})}\BibitemShut
  {NoStop}%
\bibitem [{\citenamefont {Mulliken}(1967)}]{ref35}%
  \BibitemOpen
  \bibfield  {author} {\bibinfo {author} {\bibfnamefont {R.~S.}\ \bibnamefont
  {Mulliken}},\ }\bibfield  {title} {\bibinfo {title} {Spectroscopy, molecular
  orbitals, and chemical bonding},\ }\href
  {https://doi.org/10.1126/science.157.3784.13} {\bibfield  {journal} {\bibinfo
   {journal} {Science}\ }\textbf {\bibinfo {volume} {157}},\ \bibinfo {pages}
  {13} (\bibinfo {year} {1967})}\BibitemShut {NoStop}%
\end{thebibliography}%
\end{document}